\documentclass[useAMS,usenatbib,fleqn]{mn2e}

\newcommand{\lya}{Ly$\alpha$} 
\newcommand{\msun}{\ifmmode M_{\odot} \else M$_{\odot}$\fi}
\newcommand{\msunyr}{\ifmmode M_{\odot}.{\rm yr}^{-1} \else
M$_{\odot}$ .yr$^{-1}$\fi}
\newcommand{\kms}{\ifmmode {\rm km s}^{-1} \else km s$^{-1}$\fi}

\usepackage{times}
\usepackage{graphicx}
\usepackage{epsfig}
\usepackage{amssymb}
\usepackage{natbib}
\usepackage[latin1]{inputenc}
\graphicspath{plots/}
\usepackage{amsmath}
\usepackage{stfloats}
\usepackage{fixltx2e}
\usepackage{array}
\usepackage{wrapfig}
\usepackage{relsize}
\usepackage{color}

\definecolor{r}{rgb}{1,0,0}

\title{Modelling high redshift Lyman-alpha Emitters}
\author[T. Garel et al.]{T. Garel\ensuremath{^{\textrm{1}}},
  J. Blaizot\ensuremath{^{\textrm{1}}},
  B. Guiderdoni\ensuremath{^{\textrm{1}}},
  D. Schaerer\ensuremath{^{\textrm{2,3}}},
  A. Verhamme\ensuremath{^{\textrm{1}}} and M. Hayes\ensuremath{^{\textrm{2,3}}}\\\\
\ensuremath{^{\textrm{1}}} Centre de Recherche Astrophysique de Lyon,
Universit\'e de Lyon, Universit\'e Lyon 1, Observatoire de Lyon,\\
Ecole Normale Sup\'erieure de Lyon, CNRS, UMR 5574, 9 avenue Charles
Andr\'e, Saint Genis Laval  F-69230, France\\ \ensuremath{^{\textrm{2}}} Observatoire de Gen\`eve, Universit\'e de
Gen\`eve, 51, Ch. des Maillettes, CH-1290 Versoix, Switzerland\\ \ensuremath{^{\textrm{3}}} CNRS, IRAP, 14 Avenue E. Belin, F-31400 Toulouse, France}

\begin{document}

\date{Accepted 2012 January 20. Received 2012 January 10; in original form 2011 March 01}

\pagerange{\pageref{firstpage}--\pageref{lastpage}} \pubyear{2012}

\maketitle
\label{firstpage}

\begin{abstract}
We present a new model for high redshift Lyman-Alpha Emitters (LAEs)
in the cosmological context which takes into account the resonant scattering of Ly$\alpha$
photons through expanding gas. The GALICS semi-analytic model provides
us with the physical properties of a large sample of high redshift
galaxies. We implement, in post processing, a gas outflow model for each galaxy based on simple scaling arguments. The coupling with a library of numerical experiments of Ly$\alpha$ transfer through expanding (or static) dusty shells of gas allows us to derive the Ly$\alpha$ escape fraction and profile of each galaxy. Results obtained with this new approach are compared with simpler models often used in the literature.

The predicted distribution of Ly$\alpha$ photons escape fraction shows that galaxies with a low star formation rate have
a $f_{\rm esc}$ of the order of unity, suggesting that, for those objects, Ly$\alpha$ may be used to trace the star formation rate assuming a given conversion law. In galaxies forming stars intensely, the
escape fraction spans the whole range from 0 to 1. 
The model is able to get a good match to the UV and Ly$\alpha$
luminosity function data at $3 < z < 5$.
We find that we are in good agreement with both the
bright Ly$\alpha$  data and the faint LAE population observed by \citet{rauch08}
at z $=3$ whereas a simpler \textit{constant Ly$\alpha$ escape
fraction} model fails
to do so.
Most of the Ly$\alpha$ profiles of our LAEs are redshifted by the diffusion in the expanding gas which suppresses IGM absorption and scattering.
The bulk of the observed Ly$\alpha$ equivalent width distribution is recovered by our model, but we fail to obtain the very large values sometimes detected.
Our predictions for stellar masses and UV LFs of LAEs show a satisfactory agreement with observational estimates.
The UV-brightest galaxies are found to show only low Ly$\alpha$
equivalent widths in our model, as it is reported by many observations
of high redshift LAEs. We interpret this effect as the joint consequence of old stellar
populations hosted by UV-bright galaxies, and high H{\sc i} column
densities that we predict for these objects, which quench preferentially resonant
Ly$\alpha$ photons via dust extinction.
\end{abstract}


\section{Introduction}

\begin{table*}
{\small
\vspace{1mm}
\begin{tabular}{|c|c|c|c|c|c|c|c|} \hline
Article & Model & Ly$\alpha$ model & Ly$\alpha$ LF & UV LFs of LAEs &
UV LFs & IGM & $\sigma_8$ \\ \hline 
\citet{ledelliou06} & SAM (GALFORM) & $f_{\rm esc} =$ const. & yes & no &
no & no & 0.93 \\ 
\citet{mao07} & ST & $f_{\rm esc} = {\rm f}_{\rm IGM} \times e^{\rm -{\rm A}_{\lambda}/1.08}$
& yes & no & yes & yes & 0.80 \\
\citet{koba07,koba10} & SAM (Mitaka) & $f_{\rm esc} =$
const./screen/slab & yes & yes & yes & yes & 0.90\\ 
\citet{nag} & GADGET2 & $f_{\rm esc} =$ const./Duty cycle & yes & yes &
yes & yes & 0.90 \\ 
\citet{tilvi09} & GADGET2 & $f_{\rm esc} = 1$/Duty cycle & yes & no & no & no &
0.82\\ 
\citet{samui09} & PS-ST & $f_{\rm esc} =$ const./Duty cycle & yes &
yes & yes & no & 0.80 \\ 
\citet{zheng10} & PMM N body & RT in IGM (no dust) & yes &
yes & yes & yes & 0.82\\
\citet{dayal08} & GADGET2 & $f_{\rm esc} =$ exp(-$\tau_{\rm IGM}$)
$\times$ const. & yes & yes & no & yes & 0.82\\ 
this paper & SAM (GALICS) & $f_{\rm esc} =$ RT & yes &
yes & yes & yes & 0.76\\ \hline
\end{tabular}}
\caption{Non-exhaustive summary of existing Ly$\alpha$ cosmological
models in the litterature. {\bf SAM}: Semi-analytic model. {\bf
PS}: Press-Schechter formalism. {\bf ST}: Sheth-Tormen formalism.
{\bf PMM}: Particle Multi Mesh. {\bf RT:} Radiation transfer}
\label{summary}
\end{table*}

High-redshift star-forming galaxies are expected to produce strong
Ly$\alpha$ emission lines \citep{Partridge,Charlot,valls}. Massive, hot stars are intense sources of
hydrogen-ionizing UV photons which turn part of the ISM gas into H{\sc ii}
regions. Ly$\alpha$ photons are produced by recombination of this gas.
Altough high-redshift Ly$\alpha$ emitting galaxies have long been sought without success, the number of detections has grown quickly during the last decade, thanks to narrow-band searches \citep{hu98,kud00,shima,ouch08,ouchi10,hu10}, deep spectroscopic follow-ups of UV-selected galaxies \citep{shapley,tapken}, and deep spectroscopic blind searches \citep{vanb05,rauch08}.

Although observed samples of high redshift Lyman-alpha Emitters (hereafter LAEs) have become large enough to derive statistical constraints (e.g. Ly$\alpha$ and UV luminosity functions, hereafter LF), uncertainties remain as a result of measurement errors and differences in survey detection thresholds. The physics involved in LAEs, and especially their Ly$\alpha$ escape fractions, are still poorly understood. Indeed, the travel of Ly$\alpha$ photons from their
emission regions through the galaxy and the intergalactic medium (IGM) is complicated.
The resonant nature of the Ly$\alpha$ line increases dramatically the
traveling path of the photons in the optically-thick interstellar
gas, enhancing dust absorption even in metal-poor galaxies. 
Spectroscopic studies of Ly$\alpha$ emitting galaxies
\citep{kunth98,pettini01,dawson02,shapley,tapken04,tapken06,tapken}
have shown that the line profile is complex, and can have many shapes
(P-Cygni, redward asymmetry, double bump). The measure of the
interstellar absorption lines with respect to Ly$\alpha$ by
\citet{shapley} suggests that gas outflows (probably triggered by supernova feedback) of neutral hydrogen take
place in those galaxies. Recent spectroscopic measurements led by \citet{mclinden} in two z $\sim 3$ LAEs support this idea. An expanding shell of gas surrounding the galaxy is often proposed as an explanation of this feature  and the general shape of the Ly$\alpha$ line \citep{tenorio99,mashesse03,verh06,dijk08}.

In the past years, there has been an intense investigation on the
properties of LAEs in the context of hierarchical galaxy 
formation, through semi-analytic or "hybrid" models, or numerical simulations 
\citep[e.g.][]{ledelliou05,ledelliou06,koba07,nag,samui09}. Although the
implementation of galaxy formation processes include 
state-of-the-art prescriptions, the modelling of the complicated 
mechanisms of Ly$\alpha$ photons transfer in galaxies, and their
escape from the galaxies, is usually very sketchy. The
authors frequently assume a \textit{constant Ly$\alpha$ escape 
fraction} model, and try to reproduce data (i.e Ly$\alpha$ luminosity functions) by adjusting the escape fraction as a free parameter ($f_{\rm esc}=0.02-0.60$ at $3
< z < 6$ according to models). This approach appears to work
in a satisfactory way, as far as it is possible to get a fit of the bright end 
of the LAE Ly$\alpha$ luminosity function. However, the deduced value of the free 
parameter $f_{\rm esc}$ is
not "explained", and these models fail to
reproduce the faint LAE population reported by \citet{rauch08} at z
$\sim 3$, down to a flux of $\sim 10^{-18}
\mbox{erg.s}^{-1}.\mbox{cm}^{-2}$. 

A \textit{duty cycle} scenario (in which only a fraction of the
galaxies are turned on as LAEs at a given time, or are able to be
detected because of selection criteria) has also been invoked to
reproduce the observed Ly$\alpha$ LF. \citet{nag} report that a stochastic scenario is favoured
compared to a \textit{constant Ly$\alpha$ escape fraction} model as a
result of the comparison with observational data. For the
\textit{duty cycle} model, they require a fraction of star forming
galaxies observable as LAEs at a given time equal to $0.07$ ($0.20$)
at z $= 3$ ($6$). \citet{samui09} fit their free parameters which contain the
Ly$\alpha$ escape fraction and the number of galaxies turned on as
LAEs, on the observed Ly$\alpha$ LFs and UV LFs of LAEs. Their
\textit{duty cycle} parameter has to vary with redshift in order to agree with the data.

\citet{tilvi09} relate the Ly$\alpha$ luminosity to the halo mass
accretion rate, and are able to reproduce the observed \lya\ LF
by fitting a single parameter, namely the product of the star-formation
efficiency and the \lya\ timescale. However, they assume that all Ly$\alpha$ photons are able to escape
their model galaxies ($f_{\rm esc}=1$), which is not consistent
with observations of LAEs and Lyman Break Galaxies (hereafter LBGs) \citep[e.g.][]{hayes10}.

More physical models, taking into account the properties of the
galaxies (assuming slab and screen-type dust attenuation),
have been investigated by \citet{koba07,koba10} and \citet{mao07}.
\citet{koba07,koba10} need two free parameters to reproduce the Ly$\alpha$ LF data over the redshift range $3 < z <
6$. \citet{mao07} reproduce the Ly$\alpha$ LFs data at z $= 4.9,
5.7$ and $6.4$, but they need to vary the IGM transmission. 

In parallel to these empirical approaches, several Ly$\alpha$ radiation transfer codes have been developed
\citep{zheng02,verh06,dijk06,hansen06,laursen07} including different
physics such as dust, gas kinematics, geometry, deuterium, etc. 
\citet{zheng10} perform Ly$\alpha$ radiative transfer through the
circumgalactic medium in a cosmological box, but they do not
incorporate dust into their model and do not resolve galaxies. \citet{laursen09} focus on a few high-resolution galaxies, but the CPU cost of such experiments does not allow one to process large samples of objects.
Indeed, carrying out Ly$\alpha$ line transfer in large simulated volumes, and with a resolution high enough to describe the ISM structure and kinematics, is out of CPU reach today. Hence, the need for simplified semi-analytic models remains. 
A non-exhaustive summary of the LAE models in the literature is given in Table \ref{summary}.

The purpose of this paper is to 
make one step further towards a more realistic semi-analytic approach. To this aim, we present a new model for Ly$\alpha$ emission from high redshift galaxies, which relies on two main ingredients. First, we use GALICS (for \textit{Galaxies in 
Cosmological Simulations}), a hybrid model of hierarchical galaxy 
formation in which galaxy formation and evolution are described as 
the post-processing of outputs of numerical simulations of a large
cosmological volume of dark matter (\citet{hatton}). Second, we use a large library of radiation transfer models \citep{schaerer11} computed with an updated version of MCLya \citep{verh06}, which describes the Ly$\alpha$ transfer through spherical expanding or static shells\footnote{Note that our model does not include Ly$\alpha$ radiative transfer through infalling gas.} of neutral gas and dust.
We implement a simple shell model in post-processing of GALICS, based on scaling arguments, to infer the shell parameters of the MCLya library for each model galaxy.

The advantage of this model with respect to \textit{constant Ly$\alpha$ escape fraction} models is that it computes the Ly$\alpha$ escape fraction of each model galaxy according to its physical properties. In addition, it
improves on \textit{screen} or \textit{slab} models by including the
resonant radiative transfer of the Ly$\alpha$ line, and by assuming a geometry and kinematics suggested by
the observations. 
With this new tool, we are able to compare our results with existing statistical data such as Ly$\alpha$ and UV LFs, Ly$\alpha$ equivalent width distributions, stellar masses and the Ando effect \citep[see][]{ando,koba10}.

The outline of the article is as follows. We describe
the GALICS galaxy formation model in Sec.\ 2, and the Ly$\alpha$ and shell
models in Sec.\ 3. In 
Sec.\ 4, we present the distributions of Ly$\alpha$ escape fractions we predict, and the Ly$\alpha$ LFs they yield. We discuss how these LFs are impacted by (i) equivalent width selections and (ii) IGM transmission. 
In Sec.\ 5, we show that our model matches most statistical constraints (Ly$\alpha$ equivalent width
distributions, UV LFs of LAEs, stellar masses and the Ando effect), and we use it to discuss their origin.
Finally, Sec.\ 6 summarizes the results and gives a brief discussion. 

\section{The GALICS hybrid model}
In the present paper, we use an updated version of the GALICS model
\citep{hatton,blaizot04}. We briefly describe the relevant details
below.

\subsection{Dark matter simulation}

We use a dark matter cosmological simulation run by the Horizon 
project\footnote{\tt{http://www.projet-horizon.fr}} using the public 
version of Gadget\footnote{\tt{http://www.mpa-garching.mpg.de/gadget/}} 
\citep{gadget2}. This simulation uses $1024^3$ particles of mass 
$m_{\rm p} \sim 8.5 \times 10^7$M$_\odot$ to describe the formation and evolution 
of dark matter (DM) structures in a comoving volume of 100${\rm h}^{-1}$Mpc 
on a side. It assumes a cosmology and initial conditions which are 
consistent with WMAP third year results \citep{spergel}, namely: 
$h = 0.73$, $\Lambda = 0.76$, $\Omega_{\rm m} = 0.24$, $\Omega_{\rm b} = 0.04$, 
and $\sigma_8 = 0.76$.

About 100 snapshots were saved to disk, regularly spaced in expansion 
factor by $\delta {\rm a} = 0.01$. We processed each of these snapshots to identify
DM haloes with a friends-of-friends (FOF) algorithm, using a linking length
${\rm b}=0.20$ and keeping only groups with more than 20 particles, i.e. more massive
than $1.7 \times 10^9$M$_\odot$. This mass resolution is sufficient for our 
present study, which adresses galaxy formation after reionization (z $< 5$),
when we expect the intergalactic medium's temperature to prevent gas 
from collapsing within dark matter haloes of lower masses \citep[e.g.][]{okamoto08}. Finally, we follow \citet{tweed09} to construct merger trees from our halo catalogs at all timesteps.

\subsection{Baryonic prescriptions} \label{sec:galics}

The version of GALICS we use here is an update from \citet{hatton} 
and \citet{cattaneo08}, with 3 major differences which are relevant 
for the present study: (i) the way galaxies get their gas, (ii) the way galaxies 
form stars, and (iii) the way we compute extinction of UV light by dust. 

{\it First}, the new paradigm that has emerged in recent years about gas
supply into high redshift galaxies \citep[e.g.][]{dekel06} has led us to
replace the classical gas cooling mechanism by filamentary accretion
of cold gas. In practice, for the redshift range which we explore here ($3 < z < 5$),
this means that galaxies accrete gas from the IGM at a rate directly 
proportional to the halo growth, with a delay set by the free-fall time
instead of the cooling time. 

{\it Second}, we use a Kennicutt-type law to model star formation. The low value
of $\sigma_8$ from WMAP third year results has led us to enhance star
formation significantly compared to the local law of \citet{kennicutt98}, in order
to fit high-redshift observations. In practice, we compute the star formation rate as 
\begin{equation}\label{eq:sfr}
\frac{{\rm SFR}}{\msunyr} = \mathlarger{\mathlarger{\mathlarger{\epsilon}}} 
\times 0.0328 \left(\frac{M_{\rm cold,comp}}{10^{11} \msun}\right)^{1.4} 
\left(\frac{{R_{\rm comp}}}{1 {\rm Mpc}}\right)^{-0.8},
\end{equation}
{and we assume a Kennicutt IMF \citep{kenn83}. $M_{\rm cold,comp}$ and
  $R_{comp}$ are respectively the mass of cold (i.e neutral) gas in
  the ISM and the radius of each galaxy
component: disc, bulge and burst \citep[see][for details]{hatton}. $\mathlarger{\mathlarger{\mathlarger{\epsilon}}}$ is the star formation efficiency parameter.

\begin{figure}
\includegraphics[width=0.48\textwidth,height=12.5cm]{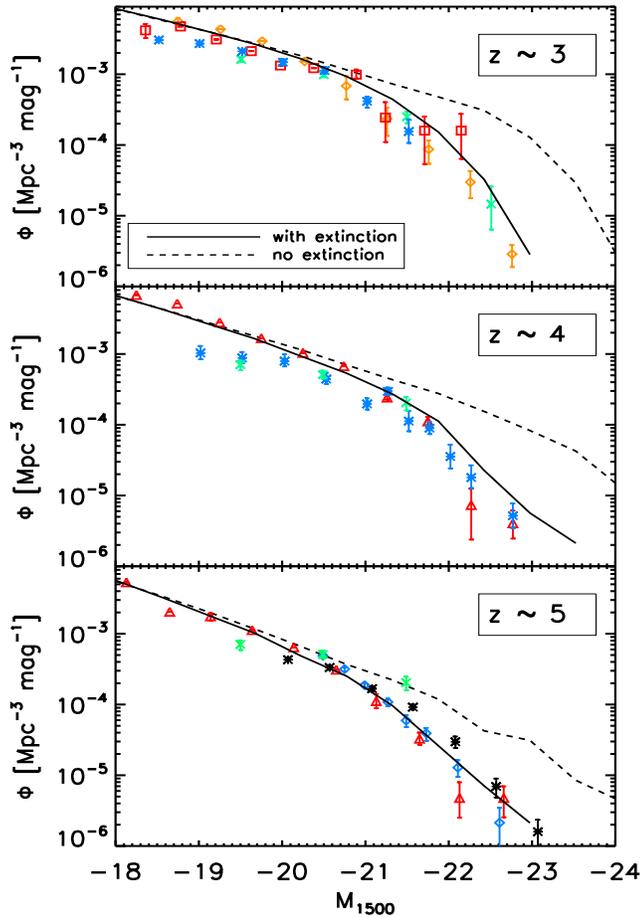}
\caption{Absolute rest-frame UV LFs (at $1500$ \AA{}) at z $\sim 3,
4$ and $5$. In each plot, the solid line refers to the UV LF after
extinction while the dashed line represents the non extinguished LF.
Data points are from \citet{reddy08} (orange diamonds),
\citet{arnouts} (red squares), \citet{sawicki} (blue asterisks),
\citet{gabasch} (green crosses), \citet{bouwens} (red triangles),
\citet{iwata} (black asterisks) and \citet{mclure} (blue diamonds).}
\label{uvlfs}
\end{figure}

{\it Third}, we now compute extinction by dust using a simple screen model, which
is consistent with our expanding shell scenario (see Sec. \ref{sec:lyamodel}), and 
we introduce a redshift dependency in the dust-to-gas ratio. In
practice, we follow \citet{hatton} and write the dust optical depth as
\begin{equation} \label{eq:tau_dust}
\tau_{\rm dust}(\lambda) = \left(\frac{A_{\lambda}}{A_V}\right)_{Z_{\odot}}
\left(\frac{Z}{Z_{\odot}}\right)^{1.35} \left(\frac{N_{\rm H}}{2.1 \times
10^{21}}\right) f(z),
\end{equation} 
where $(A_{\lambda} / A_V)_{Z_{\odot}}$ is the extinction curve for solar 
metallicity taken from \citet{mathis}, $Z$ is the metallicity of the absorbing 
gas (equal to that of the ISM), and $N_{\rm H}$ is the H{\sc i} column density. 
We compute this latter quantity with Eq. \ref{eq:column_density}, written 
for the expanding shell. It is worth noting, however, that because of our 
choice of parameters for the shell, Eq. \ref{eq:column_density} is very similar
to that used in \citet[eq. 6.3]{hatton}. The last term in Eq. \ref{eq:tau_dust}
introduces a scaling of the dust-to-gas ratio with redshift as 
$f(z) = (1+z)^{-1/2}$. This scaling is in broad agreement with obervational 
results of e.g. \citet{reddy06}, and has already been used in models, 
e.g. by \citet{kitz}. Finally, we compute the spectral energy distributions 
(SEDs) of our model galaxies with the STARDUST library \citep{devriendt}, 
as in \citet{hatton}, and extinguish them using a screen model:
\begin{equation} \label{eq:continuum_extinction}
 L_{\rm obs}(\lambda)=e^{\rm -\tau_{\rm dust}(\lambda)}L_{\rm intrinsic}(\lambda).
\end{equation}
Such a model allows us to be consistent both with the physical 
scenario we implement and with the absorption in the continuum 
found in the MCLya library (see Sec. \ref{sec:mclya_lib}).

\vskip 0.4cm

In order to adjust our model at high redshift, we want to be able to reproduce the UV LFs at $z \sim$ 3, 4, and 5. To do so, we adjust the star formation efficiency parameter $\mathlarger{\mathlarger{\mathlarger{\epsilon}}}$. $\mathlarger{\mathlarger{\mathlarger{\epsilon}}} = 1$ gives the Kennicutt law as observed at low redshifts. In the present model, we need to adopt $\mathlarger{\mathlarger{\mathlarger{\epsilon}}} = 25$ to fit the UV LFs. Although this may seem extreme, some theoritical works suggest that indeed star formation is a more violent process at high redshifts \citep{somerville01}. On the observational side, there are quite few estimates of the star formation efficiency at high redshift. \citet{baker04} measured the SFR and molecular gas density in a z $=$ 3 LBG and found that the relation between them agrees with the $\mathlarger{\mathlarger{\mathlarger{\epsilon}}} = 1$ Kennicutt law. However, using their molecular gas density measurement at $1\sigma$ can yield $\mathlarger{\mathlarger{\mathlarger{\epsilon}}} = 5$. With a recent WMAP-5 cosmology simulation, we find that GALICS can reproduce the UV LF between z $=$ 3 and 5 with a star formation efficiency $\mathlarger{\mathlarger{\mathlarger{\epsilon}}}$ of only 5. We have checked that it has very little impact on the statistical properties of high-redshift galaxies in our model. More importantly, the results of the Ly$\alpha$ model remain fully consistent with those presented in the present article. Therefore, we think that, even if it may appear as a strong deviation from local values, the high-redshift star formation efficiency we have used is not a serious problem, and can be decreased with simulation runs with an updated cosmology. These results will be presented in a next paper (Garel et al., in prep). 
Also, and perhaps more importantly, the idea of the present work is to use GALICS as a framework 
to explore the implications of our model for Ly$\alpha$ emission. In this 
prospect, it is only important for us here to have a model which reproduces 
somehow galaxy properties at high redshift.

In Figure \ref{uvlfs}, we show the rest-frame UV LFs in a
filter centered at $1500$ \AA{}, at $z \sim$ 3, 4, and 5, with $\mathlarger{\mathlarger{\mathlarger{\epsilon}}} = 25$. In each panel, 
the solid line shows our predictions (including the effect of dust) and gives a good match to the observational data. The dashed line shows our predictions prior to extinction. The strong attenuation
($\sim 1$ mag) we find at the bright end corresponds to the lower limit suggested by 
the analysis of LBGs \citep{pettini98,steidel99,blaizot04}. 

We can now turn 
to investigating the Ly$\alpha$ properties of our high-redshift model
galaxies.

\section{Ly${\alpha}$ model} \label{sec:lyamodel}

One can write the Ly$\alpha$ luminosity $L_{\rm Ly\alpha}$ of a galaxy as
\begin{equation} \label{eq:Llya}
L_{\rm Ly\alpha} = L_{\rm Ly\alpha}^{\rm intr} \times  f_{\rm esc},
\end{equation}
where $L_{\rm Ly\alpha}^{\rm intr}$ is the {\it intrinsic} Ly$\alpha$ luminosity, 
and $f_{\rm esc}$ is the fraction of these photons that actually escape the galaxy.
The first term is dominated by recombinations from photo-ionized gas
in H{\sc ii} regions, and we compute it in Sec. \ref{sec:intrinsic_llya}. 
The second term is the result from complex resonant radiative transfer. 
We present our model for $f_{\rm esc}$ in Sec. \ref{sec:fiducial_model},
and discuss its basic properties. In Sec. \ref{sec:other_models}, for the 
sake of discussion and comparison, we present a selection of alternative 
models found in the litterature. 

The possible attenuation of the Ly$\alpha$ line by the IGM is discussed later (cf \ref{sec:igm}).

\subsection{Intrinsic Ly$\alpha$ luminosities} \label{sec:intrinsic_llya}

We compute the production rate of hydrogen-ionizing photons $Q(H)$ by
integrating each galaxy's SED up to $912$ \AA.
We then write the intrinsic Ly$\alpha$ luminosity as:
\begin{equation} \label{eq:lya_caseB}
L_{Ly\alpha}^{\rm intr}= \frac{2}{3}Q(H)(1-f_{\rm esc}^{\rm ion})
\frac{hc}{\lambda_\alpha},
\end{equation} 
where $\lambda_\alpha = 1216$ \AA{} is the Ly$\alpha$ line center,
$f_{\rm esc}^{\rm ion}$ is the escape fraction of ionizing photons, $c$ the speed
of light, $h$ the Planck constant, and the factor $\frac{2}{3}$ comes
from the case B recombination \citep{oster89}. Throughout this paper, 
we assume that galaxies are ionization-bound so that $f_{\rm esc}^{\rm ion}=0$.

We assume the intrinsic Ly$\alpha$ line profile ($\Phi$) to 
be a Gaussian centered on $\lambda_\alpha$ and with a width given by the 
rotational velocity $v_{\rm rot}$ of the sources in the gravitational 
potential of the galaxy:
\begin{equation} \label{eq:input_line}
\Phi(\lambda) = \frac{c}{\sqrt{\pi} v_{\rm rot}
\lambda_\alpha}e^{-(\frac{c(1-\lambda/\lambda_\alpha)}{v_{\rm rot}})^2}.
\end{equation}

The intrinsic Ly$\alpha$ equivalent width ($EW^{\rm intr}_{\rm Ly\alpha}$) is simply
\begin{equation} \label{eq:intrinsic_ew}
EW^{\rm intr}_{\rm Ly\alpha} = \frac{L^{\rm intr}_{\rm Ly\alpha}}{L^{\rm intr}_{1216}},
\end{equation} 
where $L^{\rm intr}_{1216}$ is the unattenuated continuum luminosity estimated
by integrating each galaxy's SED from $1200$ \AA{} to $1230$ \AA{}.

\subsection{Fiducial radiative transfer model} \label{sec:fiducial_model}

In our model, the Ly$\alpha$ line properties are determined by
resonant scattering through a gas outflow. In practice, we compute the Ly$\alpha$ line properties for each model galaxy as a post-processing step of GALICS as follows. First, we follow \citet{verh08} and model the gas outflow as an expanding shell of neutral gas and dust. We relate the shell parameters to each model galaxy's physical properties in Section \ref{sec:shell}. Second, we use the \citet{schaerer11} numerical library to derive accurately the Ly$\alpha$ profile and escape fraction for each galaxy.

Here, we briefly present this library, and then describe the shell model we assume for each galaxy.

\subsubsection{MCLya library} \label{sec:mclya_lib}
\citet{schaerer11} have extended the work of \citet{verh08} by constructing
a library of numerical experiments in which they compute the transfer of Ly$\alpha$ 
photons from a central source through an expanding (or static) spherical, homogeneous shell
of mixed H{\sc i} and dust. In their model, a shell is described by four 
parameters: its expansion velocity $V_{\rm exp}$, its H{\sc i} column density $N_{\rm H}$, 
its dust opacity $\tau_{\rm dust}$, and the velocity dispersion of the gas within the shell $b$.
The library constructed by \citet{schaerer11} explores a wide range of these 
parameters, which we summarize in Table \ref{grid}, and consists of more than $5000$ 
models. Note that for simplicity, we have fixed one parameter ($b$) to a constant value of
$b = 20$ km.s$^{-1}$ (which corresponds to a typical gas temperature T $\sim 10^4$ K). This choice is motivated both by the fact that \citet{verh06} have 
shown this parameter to have the least impact on their results, and by the fact
that there is no clear physical way to vary this parameter for each of our galaxies. 

In each experiment, photons are emitted from the central source with frequencies 
ranging from $-6000$ to $+6000$ km.s$^{-1}$ around the Ly$\alpha$ line.

This extent, which has been chosen in \citet{schaerer11} to compute the grid of models, is almost always sufficient to cover the whole frequency range where resonant effects play a role.

For each experiment, the library contains the escape fraction and
the observed wavelength distribution of escaping Ly$\alpha$ photons as a
function of their input wavelength. Far from the line center, the library also 
predicts extinction of the continuum by dust, and gives results consistent with 
our Eq. \ref{eq:continuum_extinction}.

In very few extreme cases (less than one object out of a thousand at any redshift, corresponding to log(N$_{\rm H}$) $>$ 21.4 and $\tau_{\rm dust} >$ 2), the expanding shells produce very damped
absorption lines blueward 1216 \AA, with extended wings which can
contribute up to 25\% extra extinction at 6000 km.s${-1}$, compared to the
non-resonant prediction of Eq. \ref{eq:continuum_extinction}. In these cases, the MCLya library
does not allow us to compute accurately the Ly$\alpha$ EW (Eq. \ref{eq:obs_ew}). However, all these galaxies have a Ly$\alpha$ EW $< 0$ \AA{} and luminosity $< 10^{42}$ erg.s$^{-1}$, which is less than the selection criteria of observations we compare our results with. We have
checked that increasing or reducing by an arbitrary 30\% the EW of the
very few galaxies in such a configuration does not change our results
in any noticeable way.

From this library, we can compute an emergent spectrum for each model as:
\begin{equation} \label{eq:spec}
S(\lambda) = \sum_i [C(\lambda_i) + \Phi(\lambda_i)] \times f_{\rm esc}^i 
\times \phi_{out}^i(\lambda),
\end{equation}
where the sum extends over emission wavelengths $\lambda_i$, $C$ is the 
stellar continuum prior to extinction, $\Phi$ is the input line profile (Eq. \ref{eq:input_line}), 
$f_{\rm esc}^i$ is the fraction of photons emitted at $\lambda_i$ which escape the 
shell, and $\phi_{out}$ is their normalized wavelength distribution. Both $C$ and 
$\Phi$ are predicted from GALICS (Secs. \ref{sec:galics} and 
\ref{sec:intrinsic_llya}), and the library gives us values for $f_{\rm esc}$ and 
$\phi_{out}$ for each shell model. The full coupling with GALICS thus requires
one more step: the prediction of the shell parameters which will allow the selection
of the appropriate MCLya model for each galaxy. 

In practice, we will need to interpolate our predicted shell parameters
($V_{\rm exp}$, $N_{\rm H}$, and $\tau_{\rm dust}$) between grid points provided by
the MCLya library. The $V_{\rm exp}$ grid is interpolated linearly whereas we
use a logarithmic interpolation for $N_{\rm H}$ and $\tau_{\rm dust}$ (it is
due to the fact that $f_{\rm esc}$ values evolve rapidly with $N_{\rm H}$
and $\tau_{\rm dust}$ compared to $V_{\rm exp}$). Also, some of the parameter 
values predicted by GALICS are found to be outside the available MCLya grid, 
in which case we simply adopt the model at the correponding boundary. 

The number of these outliers is small compared to the whole sample ($\sim 6000$ over more than $1$ million ($400,000$) at z $= 3.1$ ($4.9$)). There are no objects with $V_{\rm exp} > V_{\rm exp}^{\rm grid,max}$. Objects with $\tau_{\rm dust} > \tau_{\rm dust}^{\rm grid,max}$ (a few hundreds at any redshift) are already very faint LAEs (L$_{Ly\alpha} < 10^{41} \mbox{erg.s}^{-1}$) when we attribute them the value $\tau_{\rm dust}^{\rm grid,max}$. They would be even fainter with their \textit{true} dust opacity value, and then fall below the luminosity limit we are interested in the present study. 
Galaxies displaying a shell column density higher than $N_{\rm H}^{\rm grid,max}$ are the most numerous (a few thousands at any redshift). All of them have Ly$\alpha$ luminosity $L_{Ly\alpha} < 5 \times 10^{42} \mbox{erg.s}^{-1}$ and an equivalent width less than $30$ \AA{}. Making the calculation with their \textit{real} $N_{\rm H}$ value would tend to reduce even more their escape fraction (and consequently their Ly$\alpha$ luminosity and equivalent width). We did the extreme test of setting all the Ly$\alpha$ luminosities of the outliers to zero and found that it does not affect the results and conclusions of the article.

\begin{table}

\small\addtolength{\tabcolsep}{-5pt}
\begin{minipage}[b]{0.5\linewidth}
\vspace{0.2cm}
\begin{tabular}{p{0.7in} c@{\hspace{2.3mm}} c@{\hspace{2.3mm}}
c@{\hspace{2.3mm}} c@{\hspace{2.3mm}} c@{\hspace{2.3mm}}
c@{\hspace{2.3mm}} c@{\hspace{2.3mm}} c@{\hspace{2.3mm}}
c@{\hspace{2.3mm}} c@{\hspace{2.3mm}} c@{\hspace{2.3mm}}
c@{\hspace{2.3mm}} c@{\hspace{2.3mm}} c@{\hspace{2.3mm}}}
\hline
$V_{\rm exp}$ (km.s$^{-1}$) & 0 & 20 & 50 & 100 & 150 & 200 & 250 & 300 &
400 & 500 & 600 & 700 \\
\hline
\end{tabular}
\end{minipage} 
\vspace{0.cm}\\
\begin{minipage}[b]{0.5\linewidth}
\begin{tabular}{p{0.65in} c@{\hspace{1.2mm}} c@{\hspace{1.2mm}}
c@{\hspace{1.2mm}} c@{\hspace{1.2mm}} c@{\hspace{1.2mm}}
c@{\hspace{1.2mm}} c@{\hspace{1.2mm}} c@{\hspace{1.2mm}}
c@{\hspace{1.2mm}} c@{\hspace{1.2mm}} c@{\hspace{1.2mm}}
c@{\hspace{1.2mm}} c@{\hspace{1.2mm}} c@{\hspace{1.2mm}}
c@{\hspace{1.2mm}} c@{\hspace{1.2mm}} }
log $N_{\rm H}$ & 16 & 18 & 18.5 & 19 & 19.3 & 19.6 & 19.9 & 20.2 & 20.5 &
20.8 & 21.1 & 21.4 & 21.7 \\
\hline
\end{tabular}
\end{minipage}
\vspace{0.cm}\\
\begin{minipage}[b]{0.5\linewidth}\centering
\begin{tabular}{p{0.7in} c@{\hspace{4.5mm}} c@{\hspace{4.5mm}}
c@{\hspace{4.5mm}} c@{\hspace{4.5mm}} c@{\hspace{4.5mm}}
c@{\hspace{4.5mm}} c@{\hspace{4.5mm}} c@{\hspace{4.5mm}}
c@{\hspace{4.5mm}} c@{\hspace{4.5mm}} c@{\hspace{4.5mm}}
c@{\hspace{4.5mm}} c@{\hspace{4.5mm}} }
$\tau_{\rm dust}$ & 0 & 0.001 & 0.1 & 0.2 & 0.5 & 1 & 1.5 & 2 & 3 & 4 \\
\hline
\end{tabular}
\end{minipage}
\caption{Grid of parameters used from the MCLya library of
\citet{schaerer11}, assuming $b=20$ km s$^{-1}$.}

\label{grid}
\end{table}

\subsubsection{Shell model} \label{sec:shell}

{In order to make use of the MCLya library described above, we now need to derive the shell parameters (expansion velocity, column density, and dust opacity) for each model galaxy. We do this as a post-processing step\footnote{Note that this shell model is done in post-processing, not in GALICS, so that it has no impact on the subsequent gas evolution and star formation in the GALICS run.} of the GALICS run, by using simple scaling arguments as follows.

First, we use a prescription taken from \citet{bertone} for the shell velocity \citep[see also][]{shu}:

\begin{equation}
{\rm V}_{\rm exp} = 623 \left(\frac{\mbox{SFR}}{100
{\rm M}_{\odot}.{\rm yr}^{-1}}\right)^{0.145} \mbox{km.s}^{-1},
\end{equation}
which links the speed of the outflowing gas to the SFR of the galaxy.

Second, we need to estimate the size and the gas mass of shell to describe its column density. We assume the shell radius is of the order of the disc radius R and we take R$_{\rm shell}$ = R, where ${R}\sim \lambda {R}_{\rm vir} / \sqrt{2}$, with $\lambda$ the spin parameter and $R_{\rm vir}$ the virial radius of the host halo \citep[see][for details]{hatton}.
 We have checked that integrating the amount
     of ejected gas over a few Myr typically gives a mass of the same
     order as that present in the ISM. For the sake of simplicity, we
     decide to set $M^{\rm gas}_{\rm shell} = M_{\rm cold} =
     \displaystyle\sum\limits_{\rm comp} M_{\rm cold, comp}$ (the
     total mass of cold gas in the galaxy).

We can now compute the shell H{\sc i} column density as
\begin{equation} \label{eq:column_density}
{\rm N}_{\rm H} = \frac{{\rm M}^{\rm gas}_{\rm shell}}{4 \pi \mu {\rm m}_{\rm H} {\rm R}^2} \mbox{ atoms per
cm}^2,
\end{equation}
where $m_{\rm H}$ is the hydrogen atom mass and $\mu$ is the mean particle mass 
in a fully neutral gas ($\mu=1.22$).

\begin{figure*}
\begin{center}
\includegraphics[width=0.9\textwidth,height=10cm]{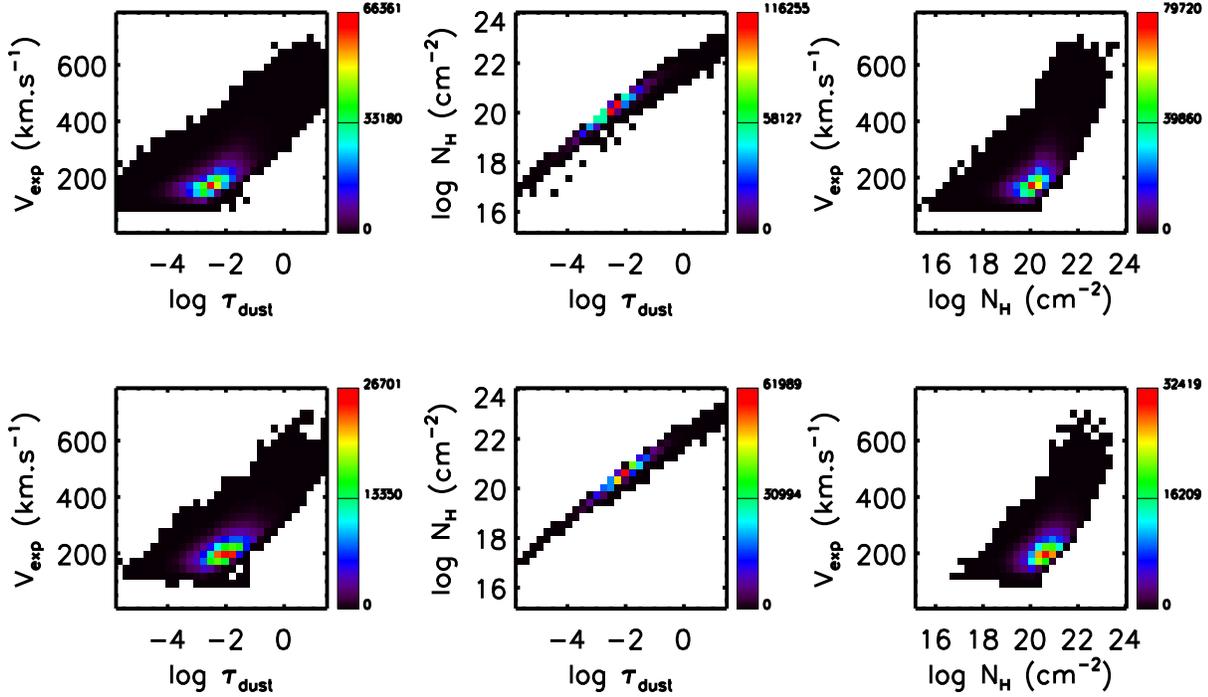}
\end{center}
\caption{Correlations between the three shell parameters at z $=
3.1$ (upper panels) and $4.9$ (lower panels) for the whole sample of
galaxies. The expansion velocity $V_{\rm exp}$ is in km.s$^{-1}$ and the
H{\sc i} column density in cm$^{-2}$. $\tau_{\rm dust}$ is the dust opacity evaluated at 1216 \AA{}. The colour code scales with the number of objects in each pixel.}
\label{param}
\end{figure*}

Finally, we compute the shell's dust optical depth at $1216$ \AA{} using 
Eq. \ref{eq:tau_dust}. Note that the models for the H{\sc i} column density and dust opacity
are identical for the Ly$\alpha$ and the UV continuum calculations. This implies
that the continuum extinction seen in the spectra from the MCLya library matches
the extinction that we apply to our galaxy SEDs. This match allows us to build full 
spectra for each model galaxy, and to measure the Ly$\alpha$ equivalent width 
directly as:
\begin{equation} \label{eq:obs_ew}
{\rm EW}_{\rm Ly\alpha} = \int \frac{{\rm S}(\lambda) - {\rm C}_{\rm ext}(\lambda)}{{\rm C}_{\rm ext}(\lambda)}
d\lambda,
\end{equation}
where ${\rm S}$ is defined in Eq. \ref{eq:spec} and ${\rm C}_{\rm ext}$ is the extinguished stellar 
continuum.

\subsubsection{Shell parameters distributions}
In Figure \ref{param}, we show our predicted distributions of the three shell
parameters at z $= 3.1$ and $4.9$ (they are similar at other
redshifts). These quantities show expected correlations. 
First, there is a tight positive correlation between $N_{\rm H}$ and $\tau_{\rm dust}$,
which directly results from our assumption that $\tau_{\rm dust} \propto N_{\rm H}$ in 
Eq. \ref{eq:tau_dust}. The small scatter accross this relation is due to metallicity.
Second, the shell velocity is a (weak) function of the SFR. Galaxies with more active
star formation have a larger reservoir of cold gas, and hence faster shells are
also those with higher H{\sc i} column densities. The linear relation
between $N_{\rm H}$ and $\tau_{\rm dust}$ is responsible for the similar
behaviour in the $V_{\rm exp}$-$N_{\rm H}$ and 
$V_{\rm exp}$-$\tau_{\rm dust}$ planes.

At all z, the H{\sc i} column density goes from $\sim 10^{16}$ to a bit
less than $10^{24}$ cm$^{-2}$. The most probable value of $N_{\rm H}$ is
$\sim 10^{20}$ ($5 \times 10^{20}$) cm$^{-2}$ at z $= 3.1$ ($4.9$). The shell velocity distributions span a whole range of values from a
few tens to $650$ km.s$^{-1}$. Most of the galaxies have $V_{\rm exp}
\sim 150-200$ km.s$^{-1}$ which is consistent with the z $= 3$ sample
of LBGs observed by \citet{shapley}. The dust opacity of the shells
ranges from ${\rm log}(\tau_{\rm dust}) = -5$ to $\sim 1.5$. The peak of the
distribution shifts from $-2.5$ at z $= 3.1$ to $-2$ at z $= 4.9$. 

\subsection{Other models for Ly$\alpha$ Emitters} \label{sec:other_models}
For discussion, we present here a selection of alternative models taken from the litterature. 

\subsubsection{Constant $f_{\rm esc}$ model}
The so-called \textit{constant Ly$\alpha$ escape fraction} model, assumes a unique escape 
fraction of Ly$\alpha$ photons for all galaxies. 
Using such a model, \citet{ledelliou06} fit the Ly$\alpha$ LF data
from z $= 3.3$ to $6.55$ with a single value $f_{\rm esc} = 0.02$. On
the other hand, \citet{nag} obtain a reasonable fit to the data by
varying $f_{\rm esc}$ with redshift, from $0.10$ at $z = 3$, to $0.15$ at $z = 6$. 

Here, we chose a value of $f_{esc} = 0.20$, which allows us to reproduce  
intermediate luminosity counts of the Ly$\alpha$ luminosity function at $z = 3.1$. 
This is also the largest value for our model not to over-predict the bright
end of the LF.

For comparison, we also explore the extreme model in which all the
Ly$\alpha$ photons are allowed to escape the galaxies, i.e $f_{\rm esc} =
1$. In the next sections, we will refer to this model as the
\textit{no extinction} model.

\subsubsection{Screen model}
In the \textit{screen} model, the fraction of Ly$\alpha$ photons
that escape the galaxy is given by 
\begin{equation}
f_{\rm esc} = e^{\rm -\tau_{\rm dust}},
\end{equation}
where $\tau_{\rm dust}$ is the dust opacity of the shell. This means that
the Ly$\alpha$ line is treated as a normal (non-resonant) radiation,
Ly$\alpha$ photons see a \textit{screen} of gas mixed with dust along 
their path. A similar model has been investigated by \citet{koba07} and
\citet{mao07} but these authors introduced an additional (free) parameter to reproduce the Ly$\alpha$ LF data.

\subsubsection{Slab model}
The \textit{slab} model \citep{koba07}, in which the escape
fraction is: 
\begin{equation}
f_{\rm esc} = \frac{(1-e^{\rm -\tau_{\rm dust}})}{\tau_{\rm dust}},
\end{equation}  
is similar to the \textit{screen} model, except that it assumes sources
are no longer behind a screen, but uniformly distributed within a slab of
gas mixed with dust. Again, and in contrast
    with us, \citet{koba07,koba10} multiplied the above $f_{\rm esc}$
    with a constant escape fraction f$_0$. These authors  specify that
    this constant parameter f$_0$ takes into account the resonant
    scattering effect of Ly$\alpha$ photons, the escape of ionizing
    photons and the IGM transmission. 

\section{Predicted Ly$\alpha$ escape fractions and Ly$\alpha$ Luminosity
Functions}
\label{sec:lyaLF}

One of the strengths of our fiducial model is that it predicts the
Ly$\alpha$ escape fraction of each individual galaxy, as a function
of its physical properties. In this section, we first discuss our predicted 
Ly$\alpha$ escape fraction distribution. Then, we compare our predicted 
Ly$\alpha$ LFs to observational estimates. We continue with discussions
on the equivalent width selection effects and IGM attenuation.

\subsection{Distribution of Ly$\alpha$ escape fractions}

In Figure \ref{dist_fesc}, we show the distribution of $f_{\rm esc}$ for
galaxies in different SFR bins, at z $= 3.1$ (thick curves) and z $=
4.9$ (thin curves). 

A first point illustrated by Figure \ref{dist_fesc} is that our model predicts a very 
strong variation of the escape fraction distribution with star formation rate 
(or, equivalently, with stellar mass).  We see that galaxies with high SFRs have a rather 
uniform $f_{\rm esc}$ distribution (solid black curves), while low-SFR objects let 
almost all Ly$\alpha$ photons escape (dashed green curves). The main quantity 
responsible for the flat distribution of the escape fraction for high-SFR galaxies
is dust opacity. Galaxies with SFR $> 20 {\rm M}_{\odot}.{\rm yr}^{-1}$ span a $\tau_{\rm dust}$ 
range going from $10^{-2}$ to more than $10$, as a consequence of their 
different star formation and merging histories.  Low-SFR objects contain little 
metal and H{\sc i} gas. Consequently, their optical thicknesses are low, and their escape fractions high.

We find that the average (median) escape fraction for galaxies with 
SFR $> 10$M$_\odot$.yr$^{-1}$ is 21\% (8\%). This compares nicely to the value
of 20\% we used to fit our \textit{constant Ly$\alpha$ escape fraction} model at intermediate Ly$\alpha$ luminosity ($10^{42} < L_{\rm Ly\alpha} < 10^{43}$erg.s$^{-1}$).

A second point we wish to make from Figure \ref{dist_fesc} is that the distribution of 
escape fractions, in a given SFR bin, remains almost constant with redshift.
The fraction of galaxies per SFR bin does not change significatively between z $= 3$ and $5$, because, from Eq. \ref{eq:sfr}, the variations (that is, a decrease with increasing redshift) of cold gas mass and disc radius balance one another. In a given SFR bin, the values of H{\sc i} column density and dust opacity (Eq. \ref{eq:column_density} and \ref{eq:tau_dust}) remain rather similar over this redshift interval, as a result of the co-evolution of cold gas mass, disc radius and metallicity. This yields the apparent non-redshift-evolution of Figure \ref{dist_fesc}.

\begin{figure}
\includegraphics[width=0.5\textwidth,height=8cm]{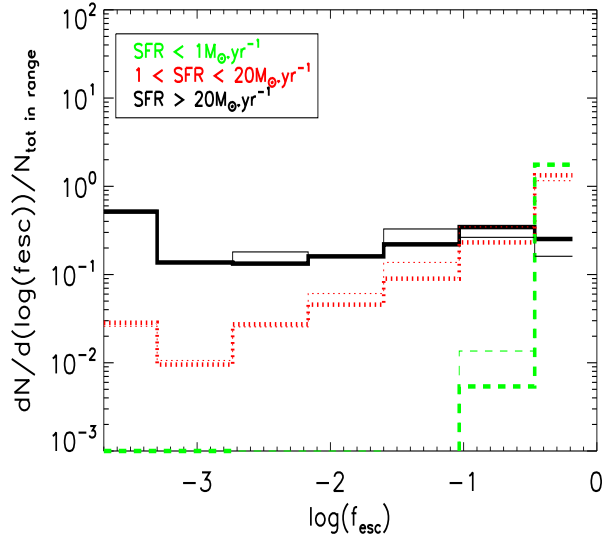}
\caption{Distribution of Ly$\alpha$ escape fraction at $z \sim 3$
(thick line) and $5$ (thin line). The black solid line refers to
galaxies having ${\rm SFR} > 20 {\rm M}_{\odot}.{\rm yr}^{-1}$, the red dotted line to
$1 < SFR < 20 {\rm M}_{\odot}.{\rm yr}^{-1}$ and the green dashed one to low-SFR objects (SFR $< 1 {\rm M}_{\odot}.{\rm yr}^{-1}$). Low-SFR galaxies have high Ly$\alpha$
escape fractions whereas in intensely star-forming objects, $f_{\rm esc}$
is distributed between $0$ and $1$.}
\label{dist_fesc}
\end{figure}

\subsection{Ly$\alpha$ luminosity functions}

\begin{figure*}
\includegraphics[width=0.9\textwidth,height=14cm]{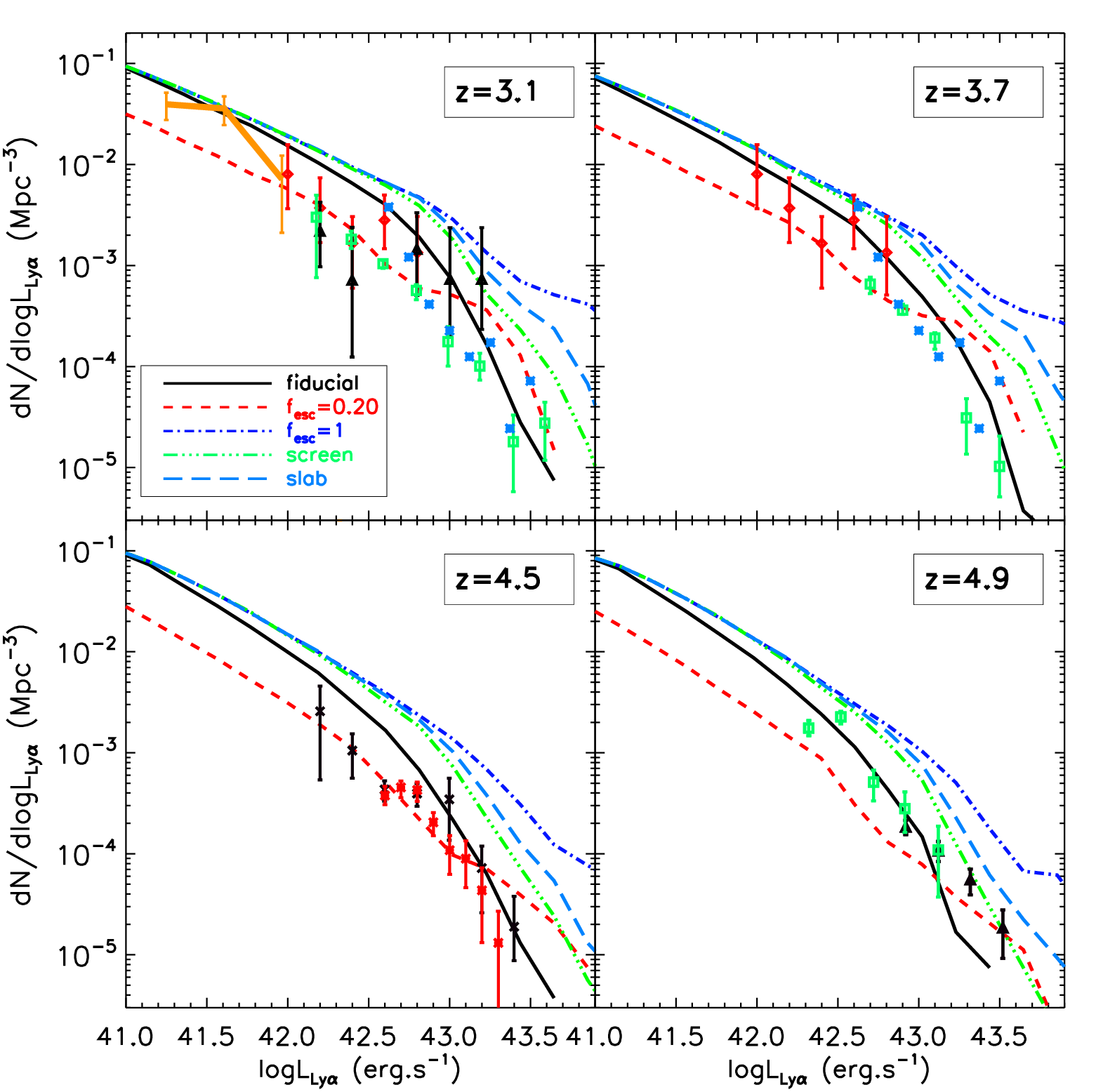}
\caption{Ly$\alpha$ LFs at z $= 3.1, 3.7, 4.5$ and $4.9$. Black solid
line: fiducial model. Red short dashed line: $f_{\rm esc} = 0.20$. Blue
long dashed line: slab model. Violet dot-dashed line: $f_{\rm esc} = 1$. Green dot-dot-dashed line: Screen model. The data points are from
\citet{vanb05} (red diamonds, $2.3 < z < 4.6$), \citet{kud00} (black
triangles, z $= 3.1$), \citet{ouch08} (green squares, z $= 3.1$, $3.7$),
\citet{blanc10} (blue asterisks, $2.8 < z < 3.8$), \citet{daws} (black crosses, z $= 4.5$),
\citet{wang09} (red asterisks, z $= 4.5$), \citet{ouch03}
(green squares, z $= 4.9$) and \citet{shioya} (black triangles, z $= 4.9$). The orange line is the observation of \citet{rauch08} ($2.67 < z < 3.75$).}
\label{lya_lfs}
\end{figure*}

In Figure \ref{lya_lfs}, we show the observed Ly$\alpha$ luminosity functions 
from $z=3.1$ to $z=4.9$, and compare them to our model (solid black curves). 
Our model shows a very satisfactory agreement with the
observational data over the whole redshift range. Interestingly, it fits
as well the bright end ($L_{\rm Ly\alpha} > 10^{42}$erg.s$^{-1}$) and
the faint LAE population observed by \citet{rauch08} at z $\sim 3$.
This is a direct result of our predicted escape fraction distribution. 
On the one hand, low-SFR 
galaxies have $f_{\rm esc} \sim 1$ due to their low dust opacities, which allows
us to reproduce the faint counts of \citet{rauch08}. On the other hand, 
high-SFR galaxies have a flat distribution of $f_{\rm esc}$, which yields the 
exponential cutoff at the bright end of the LF, as most of them have a very
low escape fraction. 

We note that, at z $=3.1$, our model agrees better with spectroscopic observations \citep{blanc10,rauch08,vanb05,kud00} than with narrow-band data from \citet{ouch08}. We will come back to 
this issue in Sec. \ref{sec:selection_effects}.

Figure \ref{lya_lfs} also shows predictions of the other models discussed in 
Sec. \ref{sec:other_models}: the blue dot-dashed (red dashed) curves 
show predictions from the $f_{\rm esc} = 1$ ($f_{\rm esc}=0.20$) model, 
the blue long-dashed (green 3-dot-dashed) curves show predictions
from the slab (screen) models. Interestingly, most models (all except 
the $f_{\rm esc}=0.20$ one) converge to the same faint-end prediction, consistent
with $f_{\rm esc} \sim 1$ for low-mass galaxies. Only our model, though, manages 
to also reproduce the bright-end, due to its resonant scattering enhancing 
Ly$\alpha$ absorption in massive, dusty, galaxies. 

At the faint end of the Ly$\alpha$ LFs where $f_{\rm esc} \sim 1$, the Ly$\alpha$ luminosity could provide information about the SFR of low mass galaxies, assuming a standard conversion law \citep{kennicutt98,furlanetto}.

\subsection{Selection effects} \label{sec:selection_effects}
Let's note that data from \citet{ouch08} (which represents the
largest sample of LAEs) around log($L_{\rm Ly\alpha}$)$ \sim 42.1-42.8$
are a bit overestimated by our model.
The theoretical Ly$\alpha$ LFs presented in Figure \ref{lya_lfs} do not contain any kind of 
selection effect. However, when selected
through narrow-band searches, as in \citet{ouch08},
observations are subject to a threshold in terms of Ly$\alpha$ equivalent width (EW).
\citet{ouch08}, especially, have a relatively high threshold at z $=3.1$ (EW$^{\rm thresh}~\sim~64$~\AA{}). Since our model is able to predict the
emergent Ly$\alpha$ EW of LAEs, we can reproduce such a selection 
and investigate its impact on LFs estimates. 

In Figure \ref{lyalf_ewcut}, we focus on the Ly$\alpha$ LF at $z=3.1$ and show 
how it varies when selecting galaxies with increasing EWs. The solid curve is the same 
as in Figure \ref{lya_lfs} (no selection), the dotted (dashed, dot-dashed) curves 
correspond to cuts at 35\AA{} (50\AA{}, 64\AA{}). Figure \ref{lyalf_ewcut} shows
that a selection on equivalent width affects the LF {\it at all luminosities}, in a rather
uniform way. Even at low luminosities ($< 10^{41}$erg.s$^{-1}$), our model galaxies have a distribution of EWs peaking at around $\sim65$\AA{}, and are thus affected 
by drastic EW cuts.

When using the threshold value of 64\AA{} quoted by \citet{ouch08} at face value, 
we find that our model under-predicts the number density of LAEs observed by these
authors (green open squares in Figure \ref{lyalf_ewcut}). Instead, we find good agreement
with their LF when applying a cut at $\sim50$\AA{}. We believe this discrepancy
has two causes: (i) our distribution of predicted EWs is perhaps centered at too low values,
and (ii) there is a rather large uncertainty in the 
estimated value of the effective EW cut from these authors' survey.
We discuss our predictions for EWs again in Sec. \ref{sec:lya_ews}.

We learn from this study that narrow-band observations may
underestimate the actual number density of LAEs at all luminosities, by a factor ranging
from 5 at the bright end to $\sim2$ at the very faint end ($L\sim10^{41}$erg.s$^{-1}$). 
Spectroscopic surveys, which are much less sensitive to EW thresholds, are more efficient to 
detect the whole sample of LAEs. Indeed, it can be seen from 
Figure \ref{lya_lfs} that most data points obtained by spectroscopy \citep{kud00,blanc10,vanb05} 
are most of the time above \citet{ouch08} observations, and in better agreement with our model predictions.
However, comparing with \citet{gronwall}'s data (who have a much lower EW limit, i.e $20$ \AA{}) does not lead to the same conclusion. \citet{gronwall}'s data (blue dashed line) are very close to those from \citet{ouch08}. Applying the $20$ \AA{} to our fiducial model does not reproduce their observed Ly$\alpha$ LF. Understanding why both \citet{ouch08} (sample of 356 objects) and \citet{gronwall} (sample of 162 objects) give a very similar luminosity function at z $=3.1$ in spite of quite different EW limits is not straightforward, given that the number of LAEs detected with EW $ < 64$ \AA{} is not negligible \citep{finkel,gronwall}. It may be a cosmic variance effect. 

In the next paragraph, we discuss what limitations arise from spectroscopic observations we have compared our model with and for which our Ly$\alpha$ LF shows a better match than with narrow-band data.

Observations of \citet{kud00} were carried out with slit spectroscopy over $\sim 50$ arcmin$^2$ so that their results may be biased by flux losses and cosmic variance. Low redshift interlopers may also have been identified as LAEs.
\citet{blanc10} apply a $20$ \AA{} equivalent width cut to remove OII emitters from their sample. According to our figure \ref{lyalf_ewcut}, such a low EW threshold should remove a small fraction of LAEs only. 
Integral field spectroscopy data from \citet{vanb05} cannot distinguish OII emitters so that their sample of LAEs may be considered as a \textit{maximal} sample. They argue that $2$ LAEs from their sample could be OII emitters. We did the test of removing those two objects which lie in the two \textit{brighter} bins of their LF. We found that our model is still in good agreement with these two points even after this correction. Nevertheless, the field of view of \citet{vanb05} is rather small ($\sim 1.4$ arcmin$^2$) and their data may suffer of cosmic variance effects. 
A more detailed discussion on pros and cons of narrow-band techniques versus integral field spectroscopy or slit spectroscopy is postponed to a future study (Garel et al., in prep).

Finally, we note that EW limits of narrow-band surveys have a decreasing effect 
with redshift (see Table \ref{thresholds}), so that the number of objects found with narrow-band and spectroscopic 
techniques should converge at higher redshifts.

\begin{figure}
\includegraphics[width=0.5\textwidth,height=8cm]{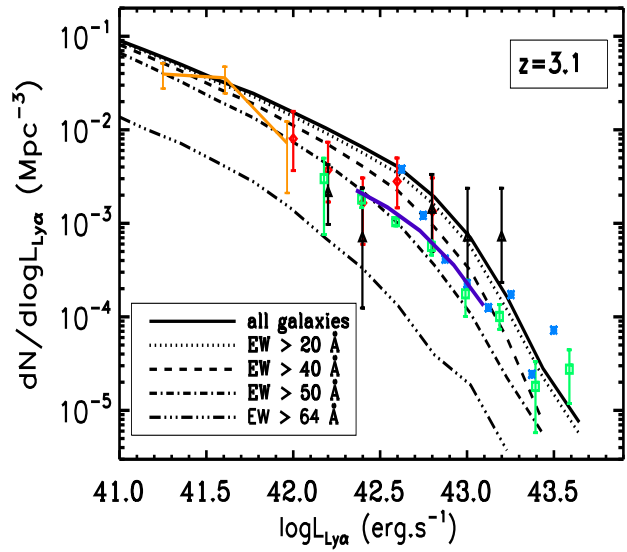}
\caption{Impact of a Ly$\alpha$ EW threshold on the Ly$\alpha$ LF at
z $= 3.1$. We plot five LFs with different cuts in Ly$\alpha$ EW.
Solid line: no cut. Dotted line: EW$^{\rm thresh} > 20$ \AA{}. Dashed
line: EW$^{\rm thresh} > 35$ \AA{}. Dot-Dot-dashed line: EW$^{\rm thresh} > 50$
\AA{}. Dot-dashed line: EW$^{\rm thresh} > 64$
\AA{}. Data points are the same as in Figure \ref{lya_lfs}. \citet{gronwall}'s data are shown as a thick violet line.}
\label{lyalf_ewcut}
\end{figure}

\subsection{Effect of the IGM} \label{sec:igm}

In the results presented so far, we have not included the effect of IGM transmission. 
However, photons shortwards $1216$\AA{} may be scattered off the line of sight by 
intergalactic hydrogen atoms. We model this effect as \citet{madau}, and define 
the IGM optical depth as:
\begin{equation}
\tau_{\rm IGM}^{\rm Ly\alpha} = 0.0036(\frac{\lambda_{\rm obs}}{\lambda_\alpha})^{3.46},
\end{equation} 
where $\lambda_{\rm obs}= (1+z)\lambda$ is the observer-frame wavelength.

We apply the IGM transmission $T_{\rm IGM}^{\rm Ly\alpha} =
e^{\rm -\tau_{\rm IGM}^{\rm Ly\alpha}}$ to the blue part of our spectra, only in the fiducial model (in which we
build the emergent Ly$\alpha$ spectra) and in the \textit{no
extinction} model (where we assume the spectrum is unchanged compared
to the Gaussian intrinsic spectrum). Other models do not produce spectra and so we discard them here. Note that if one assumes that the f$_{\rm esc} = 0.20$ model does not affect the line shape but only its amplitude, it would undergo exactly the same IGM attenuation as the \textit{no extinction model} does.

\begin{figure}
\includegraphics[width=0.5\textwidth,height=12cm]{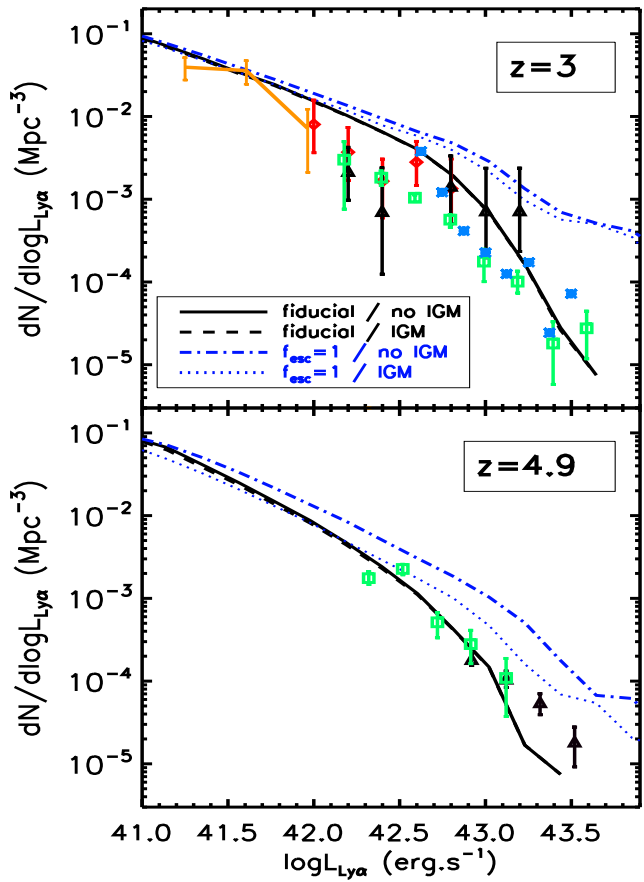}
\caption{Comparison of the Ly$\alpha$ LFs at z $= 3.1$ and $4.9$ with and without IGM transmission. The behaviour is similar at z $= 3.7$ and $4.5$. Black solid line: fiducial
model without IGM. Black dashed line: fiducial model with IGM. Violet
dot-dashed line: No extinction model without IGM. Violet dotted line:
No extinction model with IGM. Note that a horizontal shift of log(0.20) of the violet curve gives the f$_{\rm esc} = 0.20$ model in the assumption that the line shape is
unchanged by shell transfer effects. Data points are the same as in Figure
\ref{lya_lfs}.}
\label{lya_lfs_igm}
\end{figure}

In Figure \ref{lya_lfs_igm}, we show how the IGM transmission affects the 
Ly$\alpha$ LF at $3.1$ and $4.9$ only since the results at z $= 3.7$ and $4.5$ lead to the same conclusions we discuss below. 
We find that the IGM has a negligible impact on our model's Ly$\alpha$ LFs. 
This is due to the fact that, in this model, most of the galaxies' spectra have 
P-Cygni profiles, with a redward peak in emission and a deep absorption on 
the blue side. As our model for IGM transmission only applies to the blue side 
of the spectra, we indeed expect little effect from the IGM. This is probably 
a good approximation in most cases where the IGM does not produce any 
damped absorption line which could leak redwards of the Ly$\alpha$ line.
The fact that the attenuation of Ly$\alpha$ by the IGM may be relatively small or even negligible in case of outflows has already been noted by several authors, including e.g. \citet{haiman02,santos04,verh08,dijk10} and others.
In the \textit{no extinction} model, we have assumed the spectra
emerging from the galaxy are Gaussian. In this case, the
transmission through the IGM has a clear effect on the LF: it reduces luminosities by 
a factor $\sim2$ at z $= 5$. This is not enough, however,
to bring this model in agreement with the data at z $\leq 5$, which suggests that IGM
attenuation alone cannot explain the observations.

\section{Properties of Ly$\alpha$ Emitters}\label{sec:props}
We now study in more detail the properties of LAEs at $3 < z < 5$ as
predicted by our fiducial model, and we compare them to other 
available data.

\subsection{Ly$\alpha$ equivalent width} \label{sec:lya_ews}

In this section, we present the rest-frame intrinsic Ly$\alpha$ EWs
obtained from Eq. \ref{eq:intrinsic_ew}, and the rest-frame emergent (after
radiation transfer) Ly$\alpha$ EWs predicted by our fiducial model 
from Eq. \ref{eq:obs_ew}. 

In Figure \ref{ew}, we compare our predicted Ly$\alpha$ EW
distributions with observations at various redshifts (z $= 3.1, 3.7, 4.5$ and $4.9$). To perform a
reliable comparison, we apply the same criteria in terms of
Ly$\alpha$ luminosity and EW cuts as in each dataset (see Table \ref{thresholds}). 
In each panel, we show three histograms. 
The dotted green curve represents the raw distribution of intrinsic
Ly$\alpha$ EWs. The peak is at $65-70$ \AA{} at all redshifts, with
very few objects having high Ly$\alpha$ EWs ($> 100$ \AA{}). The
first reason of the deficit of high Ly$\alpha$ EWs, and of the absence of
very high Ly$\alpha$ EWs ($>200$ \AA{}) may be the absence of star formation bursts
in our GALICS galaxies. Indeed, as gas accretion is a continuous and smooth
process, the SFRs evolve smoothly and 
no galaxies show very short timescale bursts able to enhance the Ly$\alpha$ EW.
Galaxies displaying a constant SFR have rather
low Ly$\alpha$ EWs \citep{Charlot}.
Another reason for our lack of high EWs may be that we use a Kennicutt IMF. 
Considering a shallower IMF, or a higher high-mass cutoff 
could enhance the intrinsic Ly$\alpha$ EWs \citep{Charlot}. A third reason for the shallow distribution of emergent EWs could be due to 
large errors in the estimate of EWs. To take into account statistical uncertainties, 
we have convolved this distribution with a Gaussian ($\sigma = 50$ \AA{}), which yields
the green dashed curve. The choice of $50$ \AA{} is arbitrary and corresponds to the size of the bin in Figure \ref{ew} and in the Ly$\alpha$ EWs distributions commonly presented by observers. We assume that the dispersion in measurement uncertainties should not exceed this value (though it is hard to quantify). Even with this 'high' $\sigma$ value, we do not reach very high intrinsic Ly$\alpha$ EWs ($>200$ \AA{}).

We do not show the raw distribution of emergent Ly$\alpha$ EWs
obtained with our model for the sake of clarity. At z $=3.1-3.7$, it is hardly distinguishable from the intrinsic distribution. At z $=4.5-4.9$, the peak would be shifted to the $0-50$ \AA{} bin and the distribution as narrow as the raw distribution.
In Figure \ref{ew}, the solid black
line represents the distribution of emergent Ly$\alpha$ EWs convolved
with a Gaussian ($\sigma = 50$ \AA{}), as we did for the intrinsic
distribution. 
We can see that, at z $= 3.1, 3.7$ and $4.9$, the locations of the
peaks of the distributions in our predictions are in agreement with the
observations. We should note that, at z $=4.5$,
    even if the model peak matches the observed distribution from
    \citet{finkel}, it is not the case compared with \citet{daws}'s
    data. However, if we were comparing this z = $4.5$ model
    distribution with z = $4.9$ data from \citet{shioya}, we would get a good match \citep[][have nearly the same luminosity and EW detection limits so the same model can compare with these observations]{finkel,shioya,daws}. Then, we argue that it is hard to draw conclusions in that case. On the other hand, it is straightforward to conclude that all our distributions are not spread enough
compared with any data. We discuss briefly this issue.

The emergent Ly$\alpha$ EWs obtained with our fiducial model are
lower than the intrinsic ones which, as discussed above, do not reach large values and have a narrow distribution. Since the amount of dust seen by the continuum and
the Ly$\alpha$ line is the same, and given that the Ly$\alpha$ line is
resonant (and, consequently, more extinguished), it is impossible for
any galaxy to have an emergent Ly$\alpha$ EW greater than the
intrinsic one in our model. Only models with clumpy dust distributions \citep{neufeld} 
would allow $EW_{\rm Ly\alpha} > EW_{\rm Ly\alpha}^{\rm intr}$. Despite the lack of 
large EW systems, we note that our distribution reproduces a
significant fraction of observed systems, which is satisfactory. 

The reproduction of a shallow Ly$\alpha$ EW distribution with very 
large Ly$\alpha$ EWs is a puzzling issue for other models too 
\citep{samui09,dayal08}. \citet{dayal08} argue that physical effects 
such as gas kinematics, metallicity, population III stars and young stellar ages 
could spread the EW distribution, and lead to higher EW values. 
\citet{koba10} are able to retrieve the very large Ly$\alpha$ EWs thanks to 
the inclusion of both young and low-metallicity stellar populations and 
clumpy dust in their time-sequence outflow model. The value of their \textit{clumpiness parameter} (q$_{\rm d} = 0.15 =$ clumpy dust) arises from the calculation of both continuum and Ly$\alpha$ dust opacities which are computed from two different ways. 

\begin{figure*}
\includegraphics[width=0.8\textwidth,height=11cm]{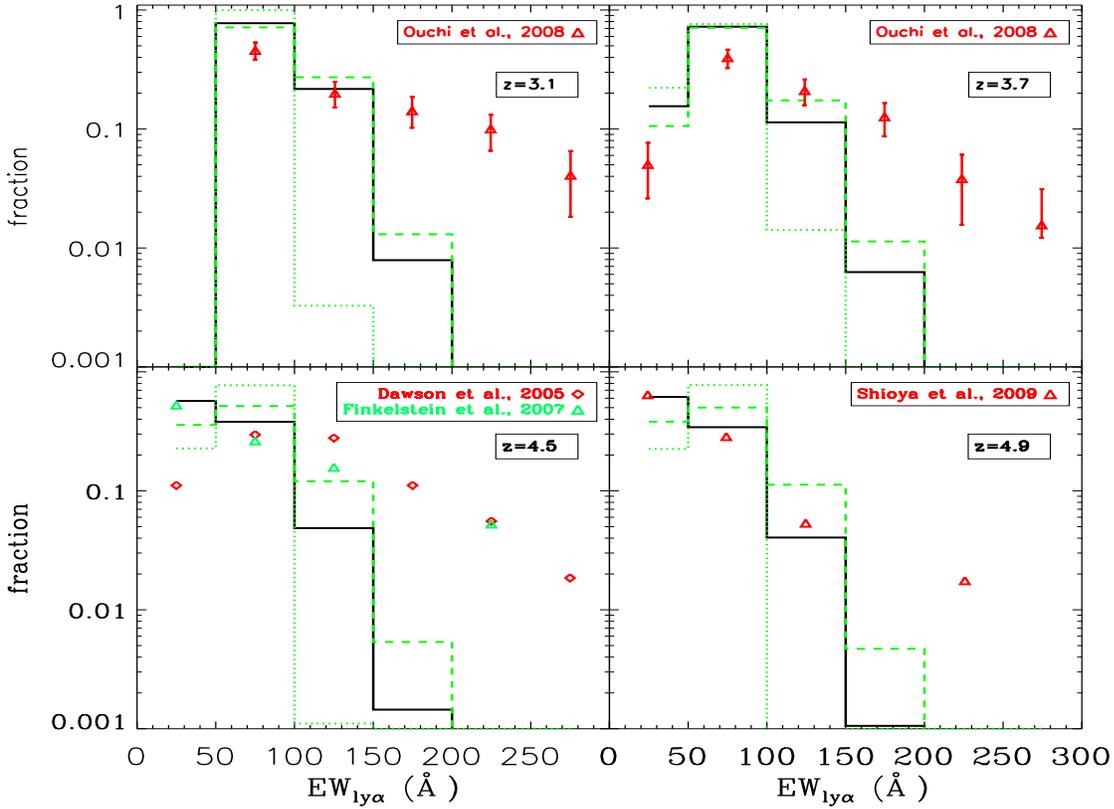}
\caption{EW distributions at z $= 3, 3.7, 4.5$ and $4.9$. Dotted
green line: Raw distribution of intrinsic Ly$\alpha$ EWs. The two
other curves have been convolved with a Gaussian ($\sigma = 50$
\AA{}) to account for statistical uncertainties. Solid black line:
Emergent Ly$\alpha$ EW distribution (fiducial model) with
convolution. Dashed green line: Intrinsic Ly$\alpha$ EW distribution
with convolution. We apply the same thresholds in terms of Ly$\alpha$
EW and luminosity as each individual set of data as summarized in
Table \ref{thresholds}. }
\label{ew}
\end{figure*}

\begin{table}
\begin{tabular}{|c|c|c|c|}
\hline
author & redshift & EW$_{\rm Ly\alpha}$ \ensuremath{^{\textrm{a}}}
(\AA{}) & $L_{\rm Ly\alpha}$ \ensuremath{^{\textrm{b}}} (erg.s$^{-1}$) \\
\hline
Ouchi et al. (2008) & z $\sim 3.1$ &  $64$ & $10^{42}$ \\
\hline
Gronwall al. (2007) & z $\sim 3.1$ &  $20$ & $1.1 \times 10^{42}$ \\
\hline
Ouchi et al. (2008) & z $\sim 3.7$ & $44$ & $4 \times 10^{42}$ \\
\hline
Dawson et al. (2007) & z $\sim 4.5$ & $14$ & $4 \times 10^{42}$ \\
\hline
Finkelstein et al. (2007) & z $\sim 4.5$ & $20$ & $4 \times 10^{42}$ \\
\hline
Wang et al. (2009) & z $\sim 4.5$ & $14$ & $3.5 \times 10^{42}$ \\
\hline
Ouchi et al. (2003) & z $\sim 4.9$ & $14$ & $7 \times 10^{41}$ \\
\hline
Shioya al. (2009) & z $\sim 4.9$ & $11$ & $3.8 \times 10^{42}$ \\
\hline
\end{tabular}
\caption{Detection limits of narrow-band surveys. a: limiting Ly$\alpha$
rest-frame EW of the survey. b: limiting Ly$\alpha$ luminosity of the
survey.}
\label{thresholds}
\end{table}

\subsection{UV Luminosity Functions of Ly$\alpha$ Emitters}

\begin{figure*}
\includegraphics[width=1.\textwidth,height=6cm]{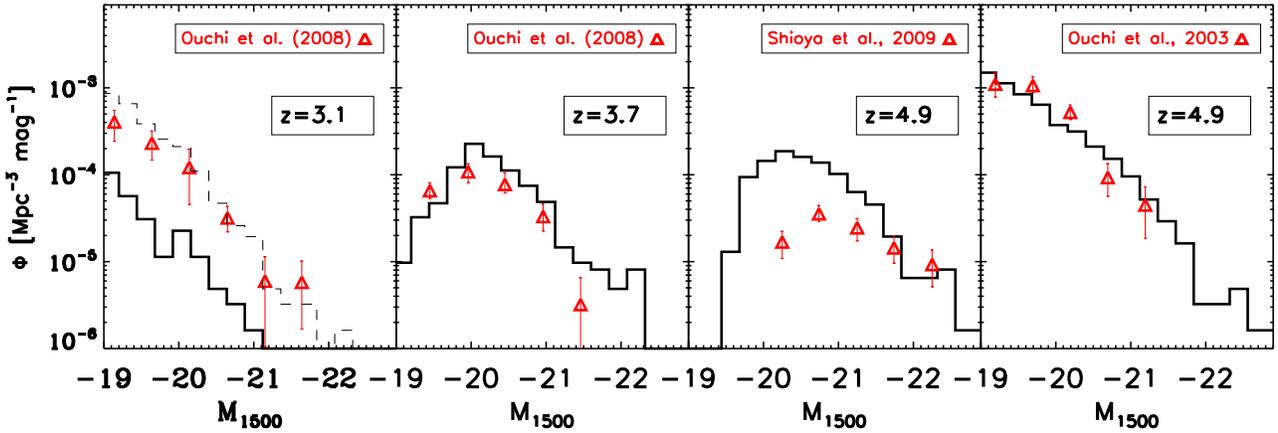}
\caption{Observed (red symbols) and predicted (black lines)
rest-frame UV LFs of LAEs at $1500$ \AA{}. 
For each LF, we apply the same cuts in Ly$\alpha$ luminosity and EW
as in the observations.
The dashed line at $z \sim 3.1$ (left panel) shows the model applying a somewhat
lower EW threshold of 50 \AA.
}
\label{uv_lae}
\end{figure*}

As noted in \citet{samui09}, only a fraction of the whole galaxy
population is detected as LAEs because of the survey limits (in
$L_{\rm Ly\alpha}$ and EW). By applying the same thresholds as in the
observations, we compute the UV LFs of LAEs at z $= 3.1, 3.7$ and $4.9$
with our fiducial model and investigate the relation between UV-selected 
galaxies (LBGs) and LAEs. 

In Figure \ref{uv_lae}, we show the UV LFs of Ly$\alpha$-selected model galaxies.
We find a rather good
agreement with observations, especially with \citet{ouch08}
at z $= 3.7$, and with \citet{ouch03} at z $= 4.9$.
However, there are two discrepancies we wish to comment on.

As already discussed with Figure \ref{lyalf_ewcut}, the EW limit of \citet{ouch08} 
at z $= 3.1$ ($64$ \AA{}) has a dramatic effect on our model, since we predict 
very few objects with large EWs. As a consequence, if we reproduce the same EW 
cut, we again find less LAEs than these authors (solid histogram in left-hand-side
panel of Figure \ref{uv_lae}). To bypass this conflict, we may lower the EW cut we apply 
to our model until we find the same number density of LAEs. We obtain this 
match at $\sim$50\AA{}, which is the value we had to apply to our modelled Ly$\alpha$ LF at z $= 3.1$ to fit the data from \citet{ouch08}. The UV LF of our model galaxies selected in this way
is plotted as the dashed curve on Figure \ref{uv_lae}. The good agreement we find now
tells us that, provided we have the same number of objects, we manage to reproduce
their UV luminosity distribution. 

For other redshifts, the EW thresholds are lower, so that our lack
of high EW is no longer a problem. However, our model does not
match z $= 4.9$ data from \citet{shioya}, and we find many more UV-faint 
objects than they do. The reason of this disagreement is unclear, especially 
given that our model agrees with data from \citet{ouch03} at the same redshift.
This suggests that observations themselves may not agree one set with 
another and that more data is needed to shed light on this issue.

From this discussion, we conclude that our model is in broad agreement 
with observed UV properties of LAEs. And we once again demonstrate the 
special care that needs to be taken to reproduce selection effects.

We may now turn the question the other way around, and ask whether 
our model reproduces the Ly$\alpha$ properties of UV-selected galaxies.
\citet{shapley} studied the Ly$\alpha$ emission of LBGs at z $= 3$. They
divided their LBG sample into four bins of Ly$\alpha$ EW and found
that $\sim$25 \% of LBGs have EWs $> 20$ \AA{} and $\sim$50 \% show
Ly$\alpha$ emission (EW $> 0$ \AA{}).
It is not straightforward to apply the LBG selection to our model galaxies, 
and even more given the complex selections inherent to spectroscopic 
followups. Instead, here, we simply apply various rest-frame UV absolute 
magnitude cuts which should roughly bracket the selection of \citet{shapley}.
With a selection limit of M$_{1500} < -21$, we find that 28 \% of
the selected LBGs have EW $> 20$ \AA{} and 69 \% display
Ly$\alpha$ emission (EW $> 0$ \AA{}) at z $= 3.1$. Varying our
selection limit, we find, for M$_{1500} < -21.5$ (M$_{1500} <
-20.5$), that 25 \% (39 \%) of the objects have EW $> 20$ \AA{},
and 74 \% (71 \%) of the selected LBGs are detected in emission.
Thus the model predicts 1.75 to 3 times less LBGs with EW $> 20$ \AA{}
than LBGs simply displaying Ly$\alpha$ emission, whereas
\citet{shapley} found a factor of two. The discrepancy with their
observations may come from the rest-frame selection instead of
apparent magnitude selection, the value of the cut, and maybe the fact that they
may have missed the detection of very faint Ly$\alpha$ lines (very
low Ly$\alpha$ EW) in their sample.  

\subsection{Stellar masses of Ly$\alpha$ Emitters}

\begin{figure*}
\includegraphics[width=0.8\textwidth,height=12cm]{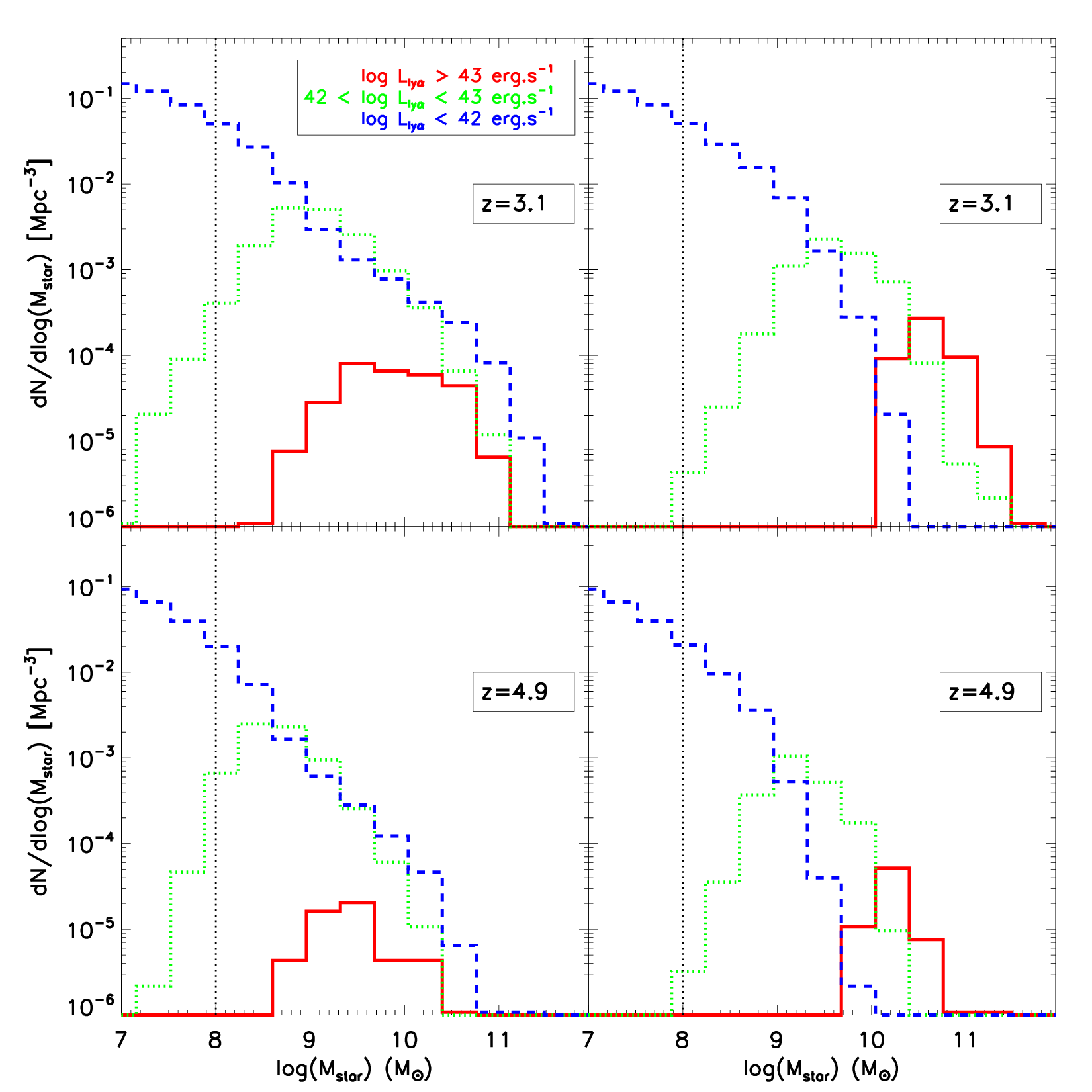}
\caption{Distribution of the stellar masses divided in three bins of
Ly$\alpha$ luminosity at z $= 3.1$ (top) and $4.9$ (bottom). In each bin, the number of objects is divided by the bin size
and the volume of the box. Solid red line:
$L_{\rm Ly\alpha} > 10^{43} \mbox{erg.s}^{-1}$. Dotted green line:
$10^{42} < L_{\rm Ly\alpha} < 10^{43} \mbox{erg.s}^{-1}$. In the left column, we show the fiducial model results. Most massive galaxies are not the brightest LAEs as a consequence of their high dust extinction. The mass ranges spanned by bright LAEs ($L_{\rm Ly\alpha} > 10^{42} \mbox{erg.s}^{-1}$, corresponding to currently observed LAEs) broadly agree with observational estimates at various redshifts. In the right column, we present the stellar mass distribution computed from the \textit{constant Ly$\alpha$ escape fraction} model ($f_{\rm esc}=0.20$), for comparison with our fiducial model. In the
\textit{constant Ly$\alpha$ escape fraction} model, the stellar mass scales with
the Ly$\alpha$ luminosity which predicts higher masses than what is
observationally derived. The mass resolution effect of the simulation starts playing a role in the stellar mass distributions at $\sim 10^8$ M$_{\odot}$ (vertical dotted line in each panel).}
\label{mstar}
\end{figure*}

Figure \ref{mstar} plots the stellar mass distributions of LAEs
divided into three Ly$\alpha$ luminosity bins at z $= 3.1$ and
$4.9$. Stellar mass distributions slowly shift to lower stellar masses by increasing the redshift. At intermediate redshifts, the results show the same behaviour as those at z $= 3.1$ and $4.9$ so we do not show them here. 

We compare the results of our fiducial model (left column) and
the $f_{\rm esc}=0.20$ model (right column). As expected,
in the latter model, brightest LAEs
($L_{\rm Ly\alpha} > 10^{43} \mbox{erg.s}^{-1}$) have higher stellar
masses, and fainter LAEs are less massive objects. It is expected
since Ly$\alpha$ luminosities scale with SFRs which is tightly 
correlated to stellar mass at these redshifts.
In our fiducial model, however, the behaviour is
slightly different. If high Ly$\alpha$ luminosity objects have medium
and rather large stellar masses (from $10^8$ to $10^{11}$ M$_{\odot}$),
the most massive objects ($>10^{11}$ M$_{\odot}$) are faint LAEs
($L_{\rm Ly\alpha} < 10^{41} \mbox{erg.s}^{-1}$). This is a consequence of
the nearly flat Ly$\alpha$ escape fraction distribution that we find for high SFR (massive)
objects (Figure \ref{dist_fesc}). For the largest fraction of LAEs which are currently
observed ($L_{\rm Ly\alpha} > 10^{42} \mbox{erg.s}^{-1}$), we predict
stellar masses ranging from $10^7$ to $10^{11}$ M$_{\odot}$.

At z $= 3.1$, \citet{gawiser06} find a mean stellar mass of
$5.10^{8}$ M$_{\odot}$ which agrees with the mean value predicted by
our fiducial model for LAEs in the range $10^{42} < L_{\rm Ly\alpha} <
10^{43} \mbox{erg.s}^{-1}$. The \textit{constant Ly$\alpha$ escape fraction}
model predicts, however, a mean value almost ten times higher for
this luminosity range.

Massive LAEs ($10^{10-11}$M$_{\odot}$) recently observed at z $= 3-4$ by \citet{ono10} have Ly$\alpha$ luminosties comprised between $\sim 10^{42}$ and $2 \times 10^{43} \mbox{erg.s}^{-1}$. Those more massive galaxies fit in the range of prediction of our model (green and red curves of the top left panel of Figure \ref{mstar}).

LAEs reported by \citet{finkel} at z $= 4.5$ have stellar masses
ranging from $2.10^{7}$ to $2.10^{9}$ M$_{\odot}$. For $L_{\rm Ly\alpha}
> 10^{42} \mbox{erg.s}^{-1}$, the fiducial model yields a mass range 
from $2.10^{7}$ to $2.10^{10}$ M$_{\odot}$, whereas the
\textit{constant Ly$\alpha$ escape fraction} model predicts higher masses. 

\citet{pirzkal07} observed LAEs with $L_{\rm Ly\alpha} > 2.10^{42}
\mbox{erg.s}^{-1}$ having $10^{7} < M_{\rm star} < 2.10^{9}$ M$_{\odot}$
at z $\sim 5$, which is rather similar to the results obtained from
the fiducial model at z $= 4.9$, and below the interval spanned by
the \textit{constant Ly$\alpha$ escape fraction} model.

Therefore, in the redshift range $3 < z < 5$, our model
gives stellar masses for bright LAEs ($L_{\rm Ly\alpha} > 10^{42} \mbox{erg.s}^{-1}$) closer to what is observed than
the \textit{constant Ly$\alpha$ escape fraction} model, and naturally recovers
the observational fact that LAEs which are currently observed are not
very massive objects. 
\subsection{Ando effect}

Many authors reported a deficit of high Ly$\alpha$ EW ($> 100$ \AA{}) in UV bright
objects (M$_{1500} < -22$) between z $= 3$ and $6$
\citep{ando,shima,ouch08,stark10}. We will refer to this effect as
the Ando effect. It has also been discussed in theoretical papers
\citep{verh08,koba10}. The reasons invoked to explain this effect are
multiple: the time-sequence of a starburst, resonant scattering in the
gas, a clumpy dust distribution and/or the age of the stellar population. We investigate this feature with our model and plot our results in Figure \ref{ando}. We
find that we recover this effect at $3 < z < 5$. Since our model does
not reproduce very accurately the observed Ly$\alpha$ EW, we do not compare
with observational data, but we only discuss the effect qualitatively. 

To see why our model predicts this lack of high Ly$\alpha$ EW in UV
bright galaxies, we show the relation between the dust-uncorrected UV
magnitude, and the intrinsic Ly$\alpha$ EW in Figure \ref{ando_intr}.
There is almost no correlation between those two quantities, except
that the highest intrinsic Ly$\alpha$ EWs come from UV faint galaxies.
It is due to the fact that UV bright objects have old stellar
populations, whereas fainter galaxies display a whole range of ages. A
fraction of the UV-faint objects are young, so that they have a high
ratio of ionizing luminosity over UV-continuum luminosity $L_{\lambda
< 912}/L_{\rm cont}$ which produces large intrinsic Ly$\alpha$ EWs. This
ratio is, on average, smaller for older, UV-brighter galaxies, so that
large intrinsic Ly$\alpha$ EWs do not exist for those objects. From
this study of the galaxy SEDs, we are able to find part of
the explanation of the absence of high Ly$\alpha$ EWs among UV-bright objects. 

Looking again at Figure \ref{ando}, we can see that this lack is more
significant for the observed Ly$\alpha$ EW (after radiative
transfer) than in the M$_{1500}^{\rm uncorr}$-EW$_{Ly\alpha}^{\rm
  intr}$ plane (Figure \ref{ando_intr}).
In our model,  H{\sc i}
column densities (and dust opacities, by construction of the dust
opacity in our model) take large values for UV-bright galaxies, as shown
by Figure \ref{mcold_uv}. We then argue that, in those galaxies,
Ly$\alpha$ photons are more extinguished than in UV-faint galaxies,
because of the resonance of the Ly$\alpha$ line in a dense, dusty medium.

As we do not reproduce the observed distribution of Ly$\alpha$ EWs at high values ($> 150$ \AA{}), we have to be prudent with our conclusions. We can wonder what would be the impact of the physical effects that we identified as a possible explanation for very large Ly$\alpha$ EWs on the Ando effect. Would clumpiness and resolved starbursts (young stellar populations) lead to high Ly$\alpha$ EW values in UV bright or faint galaxies preferentially? A possible answer can be inferred from \citet{koba10}. They find that these two effects lead to smaller (larger) Ly$\alpha$ EWs in UV brighter (fainter) galaxies. Then, the no-reproduction of large Ly$\alpha$ EWs in our model should not impact our interpretation of the Ando effect. 

Therefore, we find two main reasons to explain the Ando effect in our model: (i)
UV-bright galaxies are old, so that they do not show high intrinsic
Ly$\alpha$ EWs, and (ii) H{\sc i} column densities for UV-bright objects are larger, which leads to an enhanced
destruction of Ly$\alpha$ photons as a consequence of
radiation transfer effects, as already suggested by \citet{verh08}.

\begin{figure}
\centering
\includegraphics[width=0.4\textwidth,height=12cm]{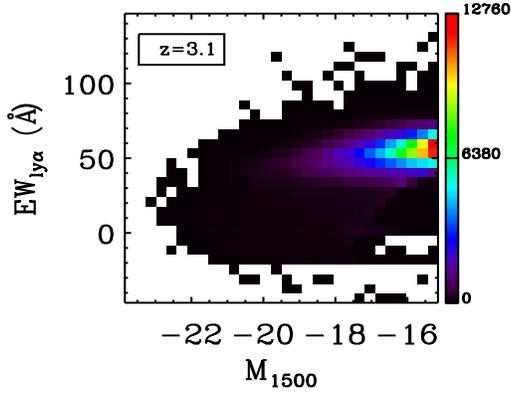}
\caption{Observed Ly$\alpha$ EW versus the UV magnitude at 1500 \AA{}
for the fiducial model at z $= 3.1$ and $4.9$. The colour of each
pixel represent the number of objects in that pixel.}
\label{ando}
\end{figure}

\begin{figure}
\includegraphics[width=0.4\textwidth,height=12cm]{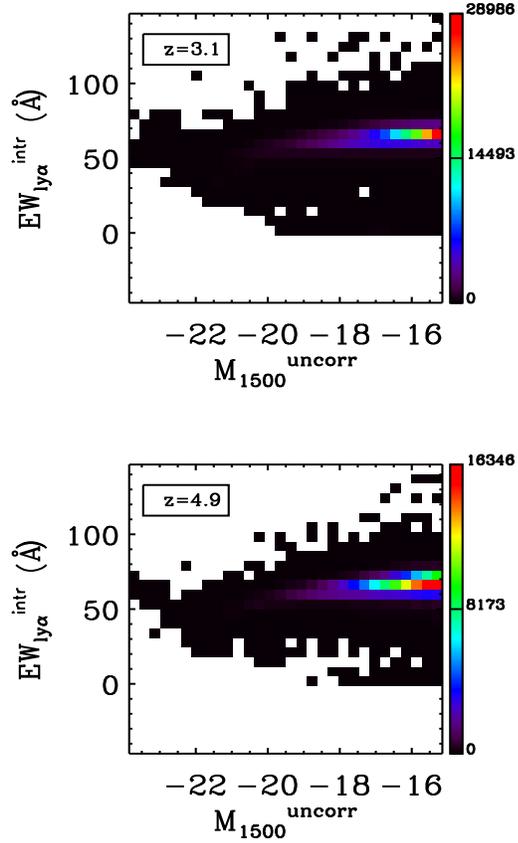}
\caption{Intrinsic Ly$\alpha$ EW versus the dust-uncorrected UV
magnitude at $1500$ \AA{} for the fiducial model at z $= 3.1$ and
$4.9$. The colour of each pixel represent the number of objects in
that pixel.}
\label{ando_intr}
\end{figure}

\begin{figure}
\includegraphics[width=0.4\textwidth,height=12cm]{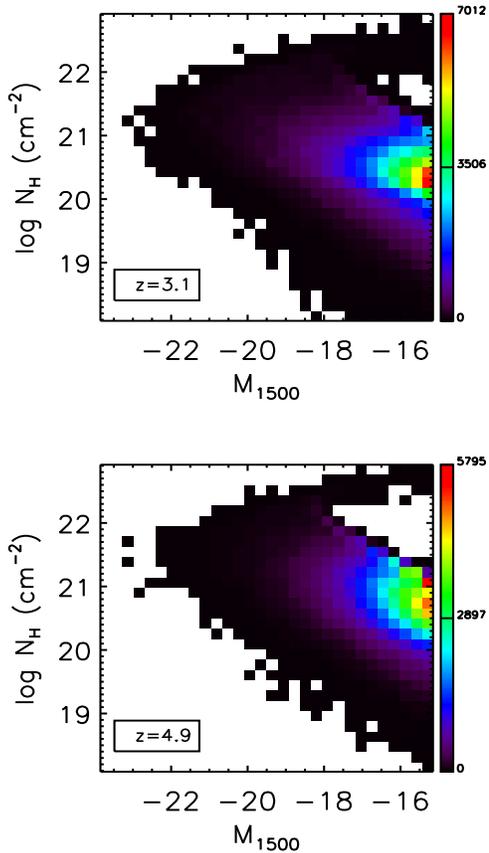}
\caption{H{\sc i} shell column density versus the UV magnitude at 1500 \AA{} for
the fiducial model at z $=$ 3.1 and 4.9. The colour of each
pixel represent the number of objects in that pixel. The no-data area (white hole) at M$_{1500}$ $\sim$ 16 and log$(N_{\rm H}) \sim$ 22 is due to the fact that (i) log$(N_{\rm H})$ is
correlated to the dust-uncorrected magnitude M$_{1500}^{\rm uncorr}$ (both proportional to the galaxy mass), and (ii) $\tau_{\rm dust}$ is proportional to N$_{\rm H}$ (Eq. \ref{eq:tau_dust}). This implies that intrinsically bright UV galaxies, in our model, have large log$(N_{\rm H})$ values and are strongly extinguished, in terms of UV magnitude, by $\sim -2.5 \times log(exp(-\tau_{\rm dust})) \sim \tau_{\rm dust} \propto N_{\rm H}$. This makes the rightward shift of large log$(N_{\rm H})$ points in this figure.}
\label{mcold_uv}
\end{figure}

\section{Discussion and conclusions} \label{sec:discussion}

In this paper, we have presented a new semi-analytic model for high
redshift LAEs. We have investigated the Ly$\alpha$ emission and transfer
processes taking into account resonant scattering effects through gas outflows. To this aim, we have coupled
the output of the GALICS semi-analytic model with results of Monte-Carlo radiation transfer runs which compute the Ly$\alpha$ transfer through static and expanding shells. We had to make a few simplifying
assumptions (central emission, sphericity and homogeneity of the
shell), and to use relations for the expanding shell that scale with the
physical properties of the galaxies as they are computed by the
semi-analytic model. 

We have run this new model on a high-resolution N-body simulation
($1024^3$ particles) of a large cosmological volume (V$=(100$h$^{-1})^3$
Mpc$^3$) of dark matter. Then, we have enough statistics for massive,
rare objects, and enough resolution for less massive objects
($M_{\rm halo}^{\rm min} = 1.70 \times 10^{9}$ M$_\odot$). In this first
paper, we aim at getting a coherent view of LAEs. We fit the UV LF at
z $= 3-5$ on a compilation of available data (Figure \ref{uvlfs}) by
adopting a high normalization of the SFR, that, in any case, scales
with gas mass as in Kennicutt's local relation. Then, we get the
following results:

\begin{itemize}
\item The Ly$\alpha$ escape fraction for each galaxy is obtained by
taking into account the resonant nature of the Ly$\alpha$ line. This
is in sharp contrast with the assumptions made in previous
semi-analytical models. The distribution of $f_{\rm esc}$ is broad, and we
see a trend with stellar masses of galaxies (Figure
\ref{dist_fesc}). Low-mass galaxies have $f_{\rm esc}$ of the order of
unity, and massive galaxies span a broad range of $f_{\rm esc}$ values. 

\item Because of this trend, the resulting Ly$\alpha$ LFs are steeper from bright to faint
luminosities than observed in simpler toy models (\textit{constant Ly$\alpha$ escape fraction}, \textit{screen} and  \textit{slab}).
 
\item Ly$\alpha$ LFs are well reproduced between z $= 3$ and $5$
(Figure \ref{lya_lfs}) without any additional free parameter in the Ly$\alpha$ model. More specifically, low-luminosity
data from \citet{rauch08} at z $\sim 3$ are reproduced, so that we
predict more faint LAEs than commonly used \textit{constant Ly$\alpha$ escape fraction} models.
 
\item We have shown that Ly$\alpha$ LFs are sensitive to Ly$\alpha$
EW cuts (Figure \ref{lyalf_ewcut}). This may explain the scatter in
the compilation of data,
since surveys (both spectroscopic and narrow-band) are
subject to different Ly$\alpha$ EW selection limits.

\item The IGM attenuation of Ly$\alpha$ photons is very weak in our
model, because the predicted Ly$\alpha$ spectra are redshifted with
respect to the Ly$\alpha$ line center, as a consequence of the scatter 
of Ly$\alpha$ photons in
the expanding shell (Figure \ref{lya_lfs_igm}). Therefore, in our model,
the Ly$\alpha$ transfer within the shell alone explains the
observed luminosities of LAEs.

\item The predicted distributions of Ly$\alpha$ EWs are narrower
than the data (Figure \ref{ew}). About 85 \% of the observed samples
have $0<{\rm EW}<150$ \AA{}, and can roughly be reproduced by the model. 
However, we predict very few objects with EW $>150$ \AA{}, whereas some are
observed. Effects that are not included in the model, like
short bursts of star formation, a top-heavy IMF, population 
III stars and/or dust clumpiness, may be the cause of such high 
Ly$\alpha$ EWs. On the other hand, even without invoking such processes, our fiducial model is able to recover roughly the bulk of the EW distribution.
 
\item The UV LFs of LAEs are in agreement with most data, with some
discrepancies (Figure \ref{uv_lae}). The scatter in the data may be
due to poorly-controlled selection criteria. 

\item We find that our predictions of the fraction of Ly$\alpha$ emitting LBGs follow
the same trend as the one found by \citet{shapley}, that is to say, $\sim 2$
times less LBGs having EW $>20$ \AA{} than LBGs having EW $>0$ \AA{}. However,
our LBG selection (in rest-frame magnitude) is somehow arbitrary
since, in this study, we do not attempt to take into account the apparent colors and magnitudes 
that are necessary to select LBGs correctly.

\item Whereas in a simple \textit{constant Ly$\alpha$ escape fraction} model,
Ly$\alpha$ luminosities scale with stellar masses, we find that most massive objects are faint LAEs (Figure
\ref{mstar}). Our predicted stellar masses for rather bright LAEs are
in correct agreement with observational estimates which find that
LAEs are intermediate-mass objects. 

\item The deficit of high Ly$\alpha$ EWs (the Ando effect) that is found
in UV-bright galaxies is well reproduced by our model (Figure
\ref{ando}). The absence of such large Ly$\alpha$ EWs comes from the
fact that H{\sc i} column densities are high for
UV-bright objects, which preferentially extinguishes Ly$\alpha$ photons, as already suggested by \citet{verh08}. Moreover, UV-bright (and consequently massive) galaxies host older stellar populations which prevent them from having high intrinsic Ly$\alpha$ EWs.
\end{itemize}

In spite of some discrepancies with specific data sets, the overall picture seems to be quite
satisfactory, given the crudeness of the assumptions. Most of the
observational constraints on high redshift LAEs are well recovered by
our model. 

Although the coupling of the semi-analytic model with Ly$\alpha$ 
radiation transfer is admittedly very crude, our global description seems 
to catch the intuitive trend according to which 
fainter galaxies, on an average, are more transparent for Ly$\alpha$ 
photons.

The hypothesis that gas outflows (with speed from a few tens to hundreds km.s$^{-1}$) are common in high redshift galaxies is well supported by observations. With such a model, we have been able to agree with many observational data and we found no need to invoke the influence of gas infalls on the Ly$\alpha$ line. Indeed, it has already been shown that it is hard to recover the redward asymetry of the Ly$\alpha$ line with models of Ly$\alpha$ radiative transfer through infalling neutral gas \citep{verh06,dijk07a}.

Obviously more refined models are still necessary, to relax some of the
assumptions, especially spherical symmetry and homogeneity of the shell. The cases
for more realistic geometries and the effect of galaxy inclination are
being investigated \citep{verh12}. 

The simulation we used in this paper has been run with initial
conditions in agreement with the WMAP 3 release, in which the $\sigma_8$
value is low. Structure growth is delayed with this low normalisation of 
the power spectrum, and fewer objects form at high redshift. This 
choice has consequences on our ability to reproduce galaxies beyond z 
$=6$, and we somehow correct this effect for lower redshifts (3 to 5) by normalizing the SFR parameter 
in order to fit the UV LFs. New
simulations with an up-to-date cosmology (WMAP 5/7), where the
derived $\sigma_8$ value is larger, can help to investigate higher redshifts 
with our approach. 

Even if the number of detections of LAEs is always increasing, the
data are still quite heterogeneous. Forthcoming LAE surveys with the Hobby-Eberly Telescope Dark Energy 
Experiment \citep[HETDEX,][]{hetdex} (z $< 3.8$; bright objects only), and the Multi Unit Spectroscopic 
Explorer \citep[MUSE,][]{bacon06} at the Very Large Telescope ($2.8 < z <6.7$) should
produce more coherent datasets. In a forthcoming paper (Garel et al., in prep.), we will present predictions for MUSE
observations  with our model.

\section{Acknowledgements}
The authors thank Roland Bacon, Steven L. Finkelstein, L\'eo Michel-Dansac, Johan Richard, Karl Joakim Rosdhal and Christian Tapken for useful
comments and discussions.
The simulation used in this work was carried out and provided by the Horizon project. 
The authors also acknowledge support from the the BINGO Project (ANR-08-BLAN-0316-01). 
DS and MH are supported by the Swiss National Science Foundation. 

We thank the anonymous referee for his/her careful reading of the manuscript, and his/her comments and suggestions that have helped the authors improve the paper.

Catalogues containing the model outputs presented in this paper can be available upon request at: thibault.garel@univ-lyon1.fr

\bibliographystyle{mn2e}
\bibliography{biblio}

\begin{thebibliography}{}

\bibitem[\protect\citeauthoryear{{Ando}, {Ohta}, {Iwata}, {Akiyama}, {Aoki} \&
  {Tamura}}{{Ando} et~al.}{2006}]{ando}
{Ando} M.,  {Ohta} K.,  {Iwata} I.,  {Akiyama} M.,  {Aoki} K.,    {Tamura} N.,
  2006, \apjl, 645, L9

\bibitem[\protect\citeauthoryear{{Arnouts}}{{Arnouts} et~al.}{2005}]{arnouts}
{Arnouts} S.,  {Schiminovich} D.,  {Ilbert} O.,  {Tresse} L.,  {Milliard} B.,
  {Treyer} M.,  {Bardelli} S.,  {Budavari} T.,  {Wyder} T.~K.,  {Zucca} E.,
  {Le F{\`e}vre} O.,  {Martin} D.~C.,  {Vettolani} G.,  {Adami} C.,
  {Arnaboldi} M.,  {Barlow} T.,  {Bianchi} L.,  {Bolzonella} M.,  {Bottini} D.,
   {Byun} Y.,  {Cappi} A.,  {Charlot} S.,  {Contini} T.,  {Donas} J.,
  {Forster} K.,  {Foucaud} S.,  {Franzetti} P.,  {Friedman} P.~G.,  {Garilli}
  B.,  {Gavignaud} I.,  {Guzzo} L.,  {Heckman} T.~M.,  {Hoopes} C.,  {Iovino}
  A.,  {Jelinsky} P.,  {Le Brun} V.,  {Lee} Y.,  {Maccagni} D.,  {Madore}
  B.~F.,  {Malina} R.,  {Marano} B.,  {Marinoni} C.,  {McCracken} H.~J.,
  {Mazure} A.,  {Meneux} B.,  {Merighi} R.,  {Morrissey} P.,  {Neff} S.,
  {Paltani} S.,  {Pell{\`o}} R.,  {Picat} J.~P.,  {Pollo} A.,  {Pozzetti} L.,
  {Radovich} M.,  {Rich} R.~M.,  {Scaramella} R.,  {Scodeggio} M.,  {Seibert}
  M.,  {Siegmund} O.,  {Small} T.,  {Szalay} A.~S.,  {Welsh} B.,  {Xu} C.~K.,
  {Zamorani} G.,    {Zanichelli} A.,  2005, \apjl, 619, L43

\bibitem[\protect\citeauthoryear{{Bacon}}{{Bacon} et~al.}{2006}]{bacon06}
{Bacon} R.,  {Bauer} S.,  {B{\"o}hm} P.,  {Boudon} D.,  {Brau-Nogue} S.,
  {Caillier} P.,  {Capoani} L.,  {Carollo} C.~M.,  {Champavert} N.,  {Contini}
  T.,  {Daguise} E.,  {Dalle} D.,  {Delabre} B.,  {Devriendt} J.,  {Dreizler}
  S.,  {Dubois} J.,  {Dupieux} M.,  {Dupin} J.,  {Emsellem} E.,  {Ferruit} P.,
  {Franx} M.,  {Gallou} G.,  {Gerssen} J.,  {Guiderdoni} B.,  {Hahn} T.,
  {Hofmann} D.,  {Jarno} A.,  {Kelz} A.,  {Koehler} C.,  {Kollatschny} W.,
  {Kosmalski} J.,  {Laurent} F.,  {Lilly} S.~J.,  {Lizon} J.,  {Loupias} M.,
  {Lynn} S.,  {Manescau} A.,  {McDermid} R.~M.,  {Monstein} C.,  {Nicklas} H.,
  {Per{\`e}s} L.,  {Pasquini} L.,  {P{\'e}contal} E.,  {P{\'e}contal-Rousset}
  A.,  {Pello} R.,  {Petit} C.,  {Picat} J.,  {Popow} E.,  {Quirrenbach} A.,
  {Reiss} R.,  {Renault} E.,  {Roth} M.,  {Schaye} J.,  {Soucail} G.,
  {Steinmetz} M.,  {Str{\"o}bele} S.,  {Stuik} R.,  {Weilbacher} P.,  {Wozniak}
  H.,    {de Zeeuw} P.~T.,  2006, The Messenger, 124, 5

\bibitem[\protect\citeauthoryear{{Baker}, {Tacconi}, {Genzel}, {Lehnert} \&
  {Lutz}}{{Baker} et~al.}{2004}]{baker04}
{Baker} A.~J.,  {Tacconi} L.~J.,  {Genzel} R.,  {Lehnert} M.~D.,    {Lutz} D.,
  2004, \apj, 604, 125

\bibitem[\protect\citeauthoryear{{Bertone}, {Stoehr} \& {White}}{{Bertone}
  et~al.}{2005}]{bertone}
{Bertone} S.,  {Stoehr} F.,    {White} S.~D.~M.,  2005, \mnras, 359, 1201

\bibitem[\protect\citeauthoryear{{Blaizot}, {Guiderdoni}, {Devriendt},
  {Bouchet}, {Hatton} \& {Stoehr}}{{Blaizot} et~al.}{2004}]{blaizot04}
{Blaizot} J.,  {Guiderdoni} B.,  {Devriendt} J.~E.~G.,  {Bouchet} F.~R.,
  {Hatton} S.~J.,    {Stoehr} F.,  2004, \mnras, 352, 571

\bibitem[\protect\citeauthoryear{{Blanc}}{{Blanc} et~al.}{2010}]{blanc10}
{Blanc} G.~A.,  {Adams} J.,  {Gebhardt} K.,  {Hill} G.~J.,  {Drory} N.,  {Hao}
  L.,  {Bender} R.,  {Ciardullo} R.,  {Finkelstein} S.~L.,  {Gawiser} E.,
  {Gronwall} C.,  {Hopp} U.,  {Jeong} D.,  {Kelzenberg} R.,  {Komatsu} E.,
  {MacQueen} P.,  {Murphy} J.~D.,  {Roth} M.~M.,  {Schneider} D.~P.,    {Tufts}
  J.,  2010, ArXiv e-prints

\bibitem[\protect\citeauthoryear{{Bouwens}, {Illingworth}, {Franx} \&
  {Ford}}{{Bouwens} et~al.}{2007}]{bouwens}
{Bouwens} R.~J.,  {Illingworth} G.~D.,  {Franx} M.,    {Ford} H.,  2007, \apj,
  670, 928

\bibitem[\protect\citeauthoryear{{Cattaneo}, {Dekel}, {Faber} \&
  {Guiderdoni}}{{Cattaneo} et~al.}{2008}]{cattaneo08}
{Cattaneo} A.,  {Dekel} A.,  {Faber} S.~M.,    {Guiderdoni} B.,  2008, \mnras,
  389, 567

\bibitem[\protect\citeauthoryear{{Charlot} \& {Fall}}{{Charlot} \&
  {Fall}}{1993}]{Charlot}
{Charlot} S.,  {Fall} S.~M.,  1993, \apj, 415, 580

\bibitem[\protect\citeauthoryear{{Dawson}, {Rhoads}, {Malhotra}, {Stern},
  {Wang}, {Dey}, {Spinrad} \& {Jannuzi}}{{Dawson} et~al.}{2007}]{daws}
{Dawson} S.,  {Rhoads} J.~E.,  {Malhotra} S.,  {Stern} D.,  {Wang} J.,  {Dey}
  A.,  {Spinrad} H.,    {Jannuzi} B.~T.,  2007, \apj, 671, 1227

\bibitem[\protect\citeauthoryear{{Dawson}, {Spinrad}, {Stern}, {Dey}, {van
  Breugel}, {de Vries} \& {Reuland}}{{Dawson} et~al.}{2002}]{dawson02}
{Dawson} S.,  {Spinrad} H.,  {Stern} D.,  {Dey} A.,  {van Breugel} W.,  {de
  Vries} W.,    {Reuland} M.,  2002, \apj, 570, 92

\bibitem[\protect\citeauthoryear{{Dayal}, {Ferrara} \& {Gallerani}}{{Dayal}
  et~al.}{2008}]{dayal08}
{Dayal} P.,  {Ferrara} A.,    {Gallerani} S.,  2008, \mnras, 389, 1683

\bibitem[\protect\citeauthoryear{{Dekel} \& {Birnboim}}{{Dekel} \&
  {Birnboim}}{2006}]{dekel06}
{Dekel} A.,  {Birnboim} Y.,  2006, \mnras, 368, 2

\bibitem[\protect\citeauthoryear{{Devriendt}, {Guiderdoni} \&
  {Sadat}}{{Devriendt} et~al.}{1999}]{devriendt}
{Devriendt} J.~E.~G.,  {Guiderdoni} B.,    {Sadat} R.,  1999, \aap, 350, 381

\bibitem[\protect\citeauthoryear{{Dijkstra}, {Haiman} \& {Spaans}}{{Dijkstra}
  et~al.}{2006}]{dijk06}
{Dijkstra} M.,  {Haiman} Z.,    {Spaans} M.,  2006, \apj, 649, 14

\bibitem[\protect\citeauthoryear{{Dijkstra}, {Lidz} \& {Wyithe}}{{Dijkstra}
  et~al.}{2007}]{dijk07a}
{Dijkstra} M.,  {Lidz} A.,    {Wyithe} J.~S.~B.,  2007, \mnras, 377, 1175

\bibitem[\protect\citeauthoryear{{Dijkstra} \& {Loeb}}{{Dijkstra} \&
  {Loeb}}{2008}]{dijk08}
{Dijkstra} M.,  {Loeb} A.,  2008, \mnras, 391, 457

\bibitem[\protect\citeauthoryear{{Dijkstra} \& {Wyithe}}{{Dijkstra} \&
  {Wyithe}}{2010}]{dijk10}
{Dijkstra} M.,  {Wyithe} J.~S.~B.,  2010, \mnras, 408, 352

\bibitem[\protect\citeauthoryear{{Finkelstein}, {Rhoads}, {Malhotra}, {Pirzkal}
  \& {Wang}}{{Finkelstein} et~al.}{2007}]{finkel}
{Finkelstein} S.~L.,  {Rhoads} J.~E.,  {Malhotra} S.,  {Pirzkal} N.,    {Wang}
  J.,  2007, \apj, 660, 1023

\bibitem[\protect\citeauthoryear{{Furlanetto}, {Schaye}, {Springel} \&
  {Hernquist}}{{Furlanetto} et~al.}{2005}]{furlanetto}
{Furlanetto} S.~R.,  {Schaye} J.,  {Springel} V.,    {Hernquist} L.,  2005,
  \apj, 622, 7

\bibitem[\protect\citeauthoryear{{Gabasch}, {Bender}, {Seitz}, {Hopp},
  {Saglia}, {Feulner}, {Snigula}, {Drory}, {Appenzeller}, {Heidt}, {Mehlert},
  {Noll}, {B{\"o}hm}, {J{\"a}ger}, {Ziegler} \& {Fricke}}{{Gabasch}
  et~al.}{2004}]{gabasch}
{Gabasch} A.,  {Bender} R.,  {Seitz} S.,  {Hopp} U.,  {Saglia} R.~P.,
  {Feulner} G.,  {Snigula} J.,  {Drory} N.,  {Appenzeller} I.,  {Heidt} J.,
  {Mehlert} D.,  {Noll} S.,  {B{\"o}hm} A.,  {J{\"a}ger} K.,  {Ziegler} B.,
  {Fricke} K.~J.,  2004, \aap, 421, 41

\bibitem[\protect\citeauthoryear{{Gawiser}}{{Gawiser} et~al.}{2006}]{gawiser06}
{Gawiser} E.,  {van Dokkum} P.~G.,  {Gronwall} C.,  {Ciardullo} R.,  {Blanc}
  G.~A.,  {Castander} F.~J.,  {Feldmeier} J.,  {Francke} H.,  {Franx} M.,
  {Haberzettl} L.,  {Herrera} D.,  {Hickey} T.,  {Infante} L.,  {Lira} P.,
  {Maza} J.,  {Quadri} R.,  {Richardson} A.,  {Schawinski} K.,  {Schirmer} M.,
  {Taylor} E.~N.,  {Treister} E.,  {Urry} C.~M.,    {Virani} S.~N.,  2006,
  \apjl, 642, L13

\bibitem[\protect\citeauthoryear{{Gronwall}, {Ciardullo}, {Hickey}, {Gawiser},
  {Feldmeier}, {van Dokkum}, {Urry}, {Herrera}, {Lehmer}, {Infante}, {Orsi},
  {Marchesini}, {Blanc}, {Francke}, {Lira} \& {Treister}}{{Gronwall}
  et~al.}{2007}]{gronwall}
{Gronwall} C.,  {Ciardullo} R.,  {Hickey} T.,  {Gawiser} E.,  {Feldmeier}
  J.~J.,  {van Dokkum} P.~G.,  {Urry} C.~M.,  {Herrera} D.,  {Lehmer} B.~D.,
  {Infante} L.,  {Orsi} A.,  {Marchesini} D.,  {Blanc} G.~A.,  {Francke} H.,
  {Lira} P.,    {Treister} E.,  2007, \apj, 667, 79

\bibitem[\protect\citeauthoryear{{Guiderdoni} \&
  {Rocca-Volmerange}}{{Guiderdoni} \& {Rocca-Volmerange}}{1987}]{guider}
{Guiderdoni} B.,  {Rocca-Volmerange} B.,  1987, \aap, 186, 1

\bibitem[\protect\citeauthoryear{{Haiman}}{{Haiman}}{2002}]{haiman02}
{Haiman} Z.,  2002, \apjl, 576, L1

\bibitem[\protect\citeauthoryear{{Hansen} \& {Oh}}{{Hansen} \&
  {Oh}}{2006}]{hansen06}
{Hansen} M.,  {Oh} S.~P.,  2006, \mnras, 367, 979

\bibitem[\protect\citeauthoryear{{Hatton}, {Devriendt}, {Ninin}, {Bouchet},
  {Guiderdoni} \& {Vibert}}{{Hatton} et~al.}{2003}]{hatton}
{Hatton} S.,  {Devriendt} J.~E.~G.,  {Ninin} S.,  {Bouchet} F.~R.,
  {Guiderdoni} B.,    {Vibert} D.,  2003, \mnras, 343, 75

\bibitem[\protect\citeauthoryear{{Hayes}, {{\"O}stlin}, {Schaerer},
  {Mas-Hesse}, {Leitherer}, {Atek}, {Kunth}, {Verhamme}, {de Barros} \&
  {Melinder}}{{Hayes} et~al.}{2010}]{hayes10}
{Hayes} M.,  {{\"O}stlin} G.,  {Schaerer} D.,  {Mas-Hesse} J.~M.,  {Leitherer}
  C.,  {Atek} H.,  {Kunth} D.,  {Verhamme} A.,  {de Barros} S.,    {Melinder}
  J.,  2010, \nat, 464, 562

\bibitem[\protect\citeauthoryear{{Hill}, {Gebhardt}, {Komatsu}, {Drory},
  {MacQueen}, {Adams}, {Blanc}, {Koehler}, {Rafal}, {Roth}, {Kelz}, {Gronwall},
  {Ciardullo} \& {Schneider}}{{Hill} et~al.}{2008}]{hetdex}
{Hill} G.~J.,  {Gebhardt} K.,  {Komatsu} E.,  {Drory} N.,  {MacQueen} P.~J.,
  {Adams} J.,  {Blanc} G.~A.,  {Koehler} R.,  {Rafal} M.,  {Roth} M.~M.,
  {Kelz} A.,  {Gronwall} C.,  {Ciardullo} R.,    {Schneider} D.~P.,  2008, in
  {T.~Kodama, T.~Yamada, \& K.~Aoki} ed., Astronomical Society of the Pacific
  Conference Series Vol.~399 of Astronomical Society of the Pacific Conference
  Series, {The Hobby-Eberly Telescope Dark Energy Experiment (HETDEX):
  Description and Early Pilot Survey Results}.
pp 115--+

\bibitem[\protect\citeauthoryear{{Hu}, {Cowie}, {Barger}, {Capak}, {Kakazu} \&
  {Trouille}}{{Hu} et~al.}{2010}]{hu10}
{Hu} E.~M.,  {Cowie} L.~L.,  {Barger} A.~J.,  {Capak} P.,  {Kakazu} Y.,
  {Trouille} L.,  2010, ArXiv e-prints

\bibitem[\protect\citeauthoryear{{Hu}, {Cowie} \& {McMahon}}{{Hu}
  et~al.}{1998}]{hu98}
{Hu} E.~M.,  {Cowie} L.~L.,    {McMahon} R.~G.,  1998, \apjl, 502, L99+

\bibitem[\protect\citeauthoryear{{Iwata}, {Ohta}, {Tamura}, {Akiyama}, {Aoki},
  {Ando}, {Kiuchi} \& {Sawicki}}{{Iwata} et~al.}{2007}]{iwata}
{Iwata} I.,  {Ohta} K.,  {Tamura} N.,  {Akiyama} M.,  {Aoki} K.,  {Ando} M.,
  {Kiuchi} G.,    {Sawicki} M.,  2007, \mnras, 376, 1557

\bibitem[\protect\citeauthoryear{{Kennicutt} Jr.}{{Kennicutt}}{1983}]{kenn83}
{Kennicutt} Jr. R.~C.,  1983, \apj, 272, 54

\bibitem[\protect\citeauthoryear{{Kennicutt}
  Jr.}{{Kennicutt}}{1998}]{kennicutt98}
{Kennicutt} Jr. R.~C.,  1998, \apj, 498, 541

\bibitem[\protect\citeauthoryear{{Kitzbichler} \& {White}}{{Kitzbichler} \&
  {White}}{2007}]{kitz}
{Kitzbichler} M.~G.,  {White} S.~D.~M.,  2007, \mnras, 376, 2

\bibitem[\protect\citeauthoryear{{Kobayashi}, {Totani} \&
  {Nagashima}}{{Kobayashi} et~al.}{2007}]{koba07}
{Kobayashi} M.~A.~R.,  {Totani} T.,    {Nagashima} M.,  2007, \apj, 670, 919

\bibitem[\protect\citeauthoryear{{Kobayashi}, {Totani} \&
  {Nagashima}}{{Kobayashi} et~al.}{2010}]{koba10}
{Kobayashi} M.~A.~R.,  {Totani} T.,    {Nagashima} M.,  2010, \apj, 708, 1119

\bibitem[\protect\citeauthoryear{{Kudritzki}, {M{\'e}ndez}, {Feldmeier},
  {Ciardullo}, {Jacoby}, {Freeman}, {Arnaboldi}, {Capaccioli}, {Gerhard} \&
  {Ford}}{{Kudritzki} et~al.}{2000}]{kud00}
{Kudritzki} R.,  {M{\'e}ndez} R.~H.,  {Feldmeier} J.~J.,  {Ciardullo} R.,
  {Jacoby} G.~H.,  {Freeman} K.~C.,  {Arnaboldi} M.,  {Capaccioli} M.,
  {Gerhard} O.,    {Ford} H.~C.,  2000, \apj, 536, 19

\bibitem[\protect\citeauthoryear{{Kunth}, {Mas-Hesse}, {Terlevich},
  {Terlevich}, {Lequeux} \& {Fall}}{{Kunth} et~al.}{1998}]{kunth98}
{Kunth} D.,  {Mas-Hesse} J.~M.,  {Terlevich} E.,  {Terlevich} R.,  {Lequeux}
  J.,    {Fall} S.~M.,  1998, \aap, 334, 11

\bibitem[\protect\citeauthoryear{{Laursen} \& {Sommer-Larsen}}{{Laursen} \&
  {Sommer-Larsen}}{2007}]{laursen07}
{Laursen} P.,  {Sommer-Larsen} J.,  2007, \apjl, 657, L69

\bibitem[\protect\citeauthoryear{{Laursen}, {Sommer-Larsen} \&
  {Andersen}}{{Laursen} et~al.}{2009}]{laursen09}
{Laursen} P.,  {Sommer-Larsen} J.,    {Andersen} A.~C.,  2009, \apj, 704, 1640

\bibitem[\protect\citeauthoryear{{Le Delliou}, {Lacey}, {Baugh}, {Guiderdoni},
  {Bacon}, {Courtois}, {Sousbie} \& {Morris}}{{Le Delliou}
  et~al.}{2005}]{ledelliou05}
{Le Delliou} M.,  {Lacey} C.,  {Baugh} C.~M.,  {Guiderdoni} B.,  {Bacon} R.,
  {Courtois} H.,  {Sousbie} T.,    {Morris} S.~L.,  2005, \mnras, 357, L11

\bibitem[\protect\citeauthoryear{{Le Delliou}, {Lacey}, {Baugh} \&
  {Morris}}{{Le Delliou} et~al.}{2006}]{ledelliou06}
{Le Delliou} M.,  {Lacey} C.~G.,  {Baugh} C.~M.,    {Morris} S.~L.,  2006,
  \mnras, 365, 712

\bibitem[\protect\citeauthoryear{{Madau}}{{Madau}}{1995}]{madau}
{Madau} P.,  1995, \apj, 441, 18

\bibitem[\protect\citeauthoryear{{Mao}, {Lapi}, {Granato}, {de Zotti} \&
  {Danese}}{{Mao} et~al.}{2007}]{mao07}
{Mao} J.,  {Lapi} A.,  {Granato} G.~L.,  {de Zotti} G.,    {Danese} L.,  2007,
  \apj, 667, 655


\bibitem[\protect\citeauthoryear{{Mas-Hesse}, {Kunth}, {Tenorio-Tagle},
  {Leitherer}, {Terlevich} \& {Terlevich}}{{Mas-Hesse}
  et~al.}{2003}]{mashesse03}
{Mas-Hesse} J.~M.,  {Kunth} D.,  {Tenorio-Tagle} G.,  {Leitherer} C.,
  {Terlevich} R.~J.,    {Terlevich} E.,  2003, \apj, 598, 858

\bibitem[\protect\citeauthoryear{{Mathis}, {Mezger} \& {Panagia}}{{Mathis}
  et~al.}{1983}]{mathis}
{Mathis} J.~S.,  {Mezger} P.~G.,    {Panagia} N.,  1983, \aap, 128, 212

\bibitem[\protect\citeauthoryear{{McLinden}, {Finkelstein}, {Rhoads},
  {Malhotra}, {Hibon} \& {Richardson}}{{McLinden} et~al.}{2011}]{mclinden}
{McLinden} E.,  {Finkelstein} S.~L.,  {Rhoads} J.~E.,  {Malhotra} S.,  {Hibon}
  P.,    {Richardson} M.,  2011, in Bulletin of the American Astronomical
  Society Vol.~43 of Bulletin of the American Astronomical Society, {First
  Spectroscopic Measurements Of [OIII] Emission From Field Lyman-alpha Selected
  Galaxies At z 3.1}.
pp 33543--+

\bibitem[\protect\citeauthoryear{{McLure}, {Cirasuolo}, {Dunlop}, {Foucaud} \&
  {Almaini}}{{McLure} et~al.}{2009}]{mclure}
{McLure} R.~J.,  {Cirasuolo} M.,  {Dunlop} J.~S.,  {Foucaud} S.,    {Almaini}
  O.,  2009, \mnras, 395, 2196

\bibitem[\protect\citeauthoryear{{Nagamine}, {Ouchi}, {Springel} \&
  {Hernquist}}{{Nagamine} et~al.}{2010}]{nag}
{Nagamine} K.,  {Ouchi} M.,  {Springel} V.,    {Hernquist} L.,  2010, \pasj,
  62, 1455

\bibitem[\protect\citeauthoryear{{Neufeld}}{{Neufeld}}{1991}]{neufeld}
{Neufeld} D.~A.,  1991, \apjl, 370, L85

\bibitem[\protect\citeauthoryear{{Okamoto}, {Gao} \& {Theuns}}{{Okamoto}
  et~al.}{2008}]{okamoto08}
{Okamoto} T.,  {Gao} L.,    {Theuns} T.,  2008, \mnras, 390, 920

\bibitem[\protect\citeauthoryear{{Ono}, {Ouchi}, {Shimasaku}, {Akiyama},
  {Dunlop}, {Farrah}, {Lee}, {McLure}, {Okamura} \& {Yoshida}}{{Ono}
  et~al.}{2010}]{ono10}
{Ono} Y.,  {Ouchi} M.,  {Shimasaku} K.,  {Akiyama} M.,  {Dunlop} J.,  {Farrah}
  D.,  {Lee} J.~C.,  {McLure} R.,  {Okamura} S.,    {Yoshida} M.,  2010,
  \mnras, 402, 1580

\bibitem[\protect\citeauthoryear{{Osterbrock}}{{Osterbrock}}{1989}]{oster89}
{Osterbrock} D.~E.,  1989, {Astrophysics of gaseous nebulae and active galactic
  nuclei}

\bibitem[\protect\citeauthoryear{{Ouchi}, {Shimasaku}, {Akiyama}, {Simpson},
  {Saito}, {Ueda}, {Furusawa}, {Sekiguchi}, {Yamada}, {Kodama}, {Kashikawa},
  {Okamura}, {Iye}, {Takata}, {Yoshida} \& {Yoshida}}{{Ouchi}
  et~al.}{2008}]{ouch08}
{Ouchi} M.,  {Shimasaku} K.,  {Akiyama} M.,  {Simpson} C.,  {Saito} T.,  {Ueda}
  Y.,  {Furusawa} H.,  {Sekiguchi} K.,  {Yamada} T.,  {Kodama} T.,  {Kashikawa}
  N.,  {Okamura} S.,  {Iye} M.,  {Takata} T.,  {Yoshida} M.,    {Yoshida} M.,
  2008, \apjs, 176, 301

\bibitem[\protect\citeauthoryear{{Ouchi}}{{Ouchi} et~al.}{2003}]{ouch03}
{Ouchi} M.,  {Shimasaku} K.,  {Furusawa} H.,  {Miyazaki} M.,  {Doi} M.,
  {Hamabe} M.,  {Hayashino} T.,  {Kimura} M.,  {Kodaira} K.,  {Komiyama} Y.,
  {Matsuda} Y.,  {Miyazaki} S.,  {Nakata} F.,  {Okamura} S.,  {Sekiguchi} M.,
  {Shioya} Y.,  {Tamura} H.,  {Taniguchi} Y.,  {Yagi} M.,    {Yasuda} N.,
  2003, \apj, 582, 60

\bibitem[\protect\citeauthoryear{{Ouchi}, {Shimasaku}, {Furusawa}, {SAITO},
  {Yoshida}, {Akiyama}, {Ono}, {Yamada}, {Ota}, {Kashikawa}, {Iye}, {Kodama},
  {Okamura}, {Simpson} \& {Yoshida}}{{Ouchi} et~al.}{2010}]{ouchi10}
{Ouchi} M.,  {Shimasaku} K.,  {Furusawa} H.,  {SAITO} T.,  {Yoshida} M.,
  {Akiyama} M.,  {Ono} Y.,  {Yamada} T.,  {Ota} K.,  {Kashikawa} N.,  {Iye} M.,
   {Kodama} T.,  {Okamura} S.,  {Simpson} C.,    {Yoshida} M.,  2010, ArXiv
  e-prints

\bibitem[\protect\citeauthoryear{{Partridge} \& {Peebles}}{{Partridge} \&
  {Peebles}}{1967}]{Partridge}
{Partridge} R.~B.,  {Peebles} P.~J.~E.,  1967, \apj, 147, 868

\bibitem[\protect\citeauthoryear{{Pettini}, {Kellogg}, {Steidel}, {Dickinson},
  {Adelberger} \& {Giavalisco}}{{Pettini} et~al.}{1998}]{pettini98}
{Pettini} M.,  {Kellogg} M.,  {Steidel} C.~C.,  {Dickinson} M.,  {Adelberger}
  K.~L.,    {Giavalisco} M.,  1998, \apj, 508, 539

\bibitem[\protect\citeauthoryear{{Pettini}, {Shapley}, {Steidel}, {Cuby},
  {Dickinson}, {Moorwood}, {Adelberger} \& {Giavalisco}}{{Pettini}
  et~al.}{2001}]{pettini01}
{Pettini} M.,  {Shapley} A.~E.,  {Steidel} C.~C.,  {Cuby} J.,  {Dickinson} M.,
  {Moorwood} A.~F.~M.,  {Adelberger} K.~L.,    {Giavalisco} M.,  2001, \apj,
  554, 981

\bibitem[\protect\citeauthoryear{{Pirzkal}, {Malhotra}, {Rhoads} \&
  {Xu}}{{Pirzkal} et~al.}{2007}]{pirzkal07}
{Pirzkal} N.,  {Malhotra} S.,  {Rhoads} J.~E.,    {Xu} C.,  2007, \apj, 667,
49

\bibitem[\protect\citeauthoryear{{Rauch}, {Haehnelt}, {Bunker}, {Becker},
  {Marleau}, {Graham}, {Cristiani}, {Jarvis}, {Lacey}, {Morris}, {Peroux},
  {R{\"o}ttgering} \& {Theuns}}{{Rauch} et~al.}{2008}]{rauch08}
{Rauch} M.,  {Haehnelt} M.,  {Bunker} A.,  {Becker} G.,  {Marleau} F.,
  {Graham} J.,  {Cristiani} S.,  {Jarvis} M.,  {Lacey} C.,  {Morris} S.,
  {Peroux} C.,  {R{\"o}ttgering} H.,    {Theuns} T.,  2008, \apj, 681, 856

\bibitem[\protect\citeauthoryear{{Reddy}, {Steidel}, {Fadda}, {Yan}, {Pettini},
  {Shapley}, {Erb} \& {Adelberger}}{{Reddy} et~al.}{2006}]{reddy06}
{Reddy} N.~A.,  {Steidel} C.~C.,  {Fadda} D.,  {Yan} L.,  {Pettini} M.,
  {Shapley} A.~E.,  {Erb} D.~K.,    {Adelberger} K.~L.,  2006, \apj, 644, 792

\bibitem[\protect\citeauthoryear{{Reddy}, {Steidel}, {Pettini}, {Adelberger},
  {Shapley}, {Erb} \& {Dickinson}}{{Reddy} et~al.}{2008}]{reddy08}
{Reddy} N.~A.,  {Steidel} C.~C.,  {Pettini} M.,  {Adelberger} K.~L.,  {Shapley}
  A.~E.,  {Erb} D.~K.,    {Dickinson} M.,  2008, \apjs, 175, 48

\bibitem[\protect\citeauthoryear{{Samui}, {Srianand} \& {Subramanian}}{{Samui}
  et~al.}{2009}]{samui09}
{Samui} S.,  {Srianand} R.,    {Subramanian} K.,  2009, \mnras, 398, 2061

\bibitem[\protect\citeauthoryear{{Santos}, {Ellis}, {Kneib}, {Richard} \&
  {Kuijken}}{{Santos} et~al.}{2004}]{santos04}
{Santos} M.~R.,  {Ellis} R.~S.,  {Kneib} J.,  {Richard} J.,    {Kuijken} K.,
  2004, \apj, 606, 683

\bibitem[\protect\citeauthoryear{{Sawicki} \& {Thompson}}{{Sawicki} \&
  {Thompson}}{2006}]{sawicki}
{Sawicki} M.,  {Thompson} D.,  2006, \apj, 648, 299

\bibitem[\protect\citeauthoryear{{Schaerer}, {Hayes}, {Verhamme} \&
  {Teyssier}}{{Schaerer} et~al.}{2011}]{schaerer11}
{Schaerer} D.,  {Hayes} M.,  {Verhamme} A.,    {Teyssier} R.,  2011, ArXiv
  e-prints


\bibitem[\protect\citeauthoryear{{Shapley}, {Steidel}, {Pettini} \&
  {Adelberger}}{{Shapley} et~al.}{2003}]{shapley}
{Shapley} A.~E.,  {Steidel} C.~C.,  {Pettini} M.,    {Adelberger} K.~L.,  2003,
  \apj, 588, 65

\bibitem[\protect\citeauthoryear{{Shimasaku}, {Kashikawa}, {Doi}, {Ly},
  {Malkan}, {Matsuda}, {Ouchi}, {Hayashino}, {Iye}, {Motohara}, {Murayama},
  {Nagao}, {Ohta}, {Okamura}, {Sasaki}, {Shioya} \& {Taniguchi}}{{Shimasaku}
  et~al.}{2006}]{shima}
{Shimasaku} K.,  {Kashikawa} N.,  {Doi} M.,  {Ly} C.,  {Malkan} M.~A.,
  {Matsuda} Y.,  {Ouchi} M.,  {Hayashino} T.,  {Iye} M.,  {Motohara} K.,
  {Murayama} T.,  {Nagao} T.,  {Ohta} K.,  {Okamura} S.,  {Sasaki} T.,
  {Shioya} Y.,    {Taniguchi} Y.,  2006, \pasj, 58, 313

\bibitem[\protect\citeauthoryear{{Shioya}}{{Shioya} et~al.}{2009}]{shioya}
{Shioya} Y.,  {Taniguchi} Y.,  {Sasaki} S.~S.,  {Nagao} T.,  {Murayama} T.,
  {Saito} T.,  {Ideue} Y.,  {Nakajima} A.,  {Matsuoka} K.,  {Trump} J.,
  {Scoville} N.~Z.,  {Sanders} D.~B.,  {Mobasher} B.,  {Aussel} H.,  {Capak}
  P.,  {Kartaltepe} J.,  {Koekemoer} A.,  {Carilli} C.,  {Ellis} R.~S.,
  {Garilli} B.,  {Giavalisco} M.,  {Kitzbichler} M.~G.,  {Impey} C.,  {LeFevre}
  O.,  {Schinnerer} E.,    {Smolcic} V.,  2009, \apj, 696, 546

\bibitem[\protect\citeauthoryear{{Shu}, {Mo} \& {Shu-DeMao}}{{Shu}
  et~al.}{2005}]{shu}
{Shu} C.,  {Mo} H.,    {Shu-DeMao} 2005, \cjaa, 5, 327

\bibitem[\protect\citeauthoryear{{Somerville}, {Primack} \&
  {Faber}}{{Somerville} et~al.}{2001}]{somerville01}
{Somerville} R.~S.,  {Primack} J.~R.,    {Faber} S.~M.,  2001, \mnras, 320, 504


\bibitem[\protect\citeauthoryear{{Spergel}}{{Spergel} et~al.}{2007}]{spergel}
{Spergel} D.~N.,  {Bean} R.,  {Dor{\'e}} O.,  {Nolta} M.~R.,  {Bennett} C.~L.,
  {Dunkley} J.,  {Hinshaw} G.,  {Jarosik} N.,  {Komatsu} E.,  {Page} L.,
  {Peiris} H.~V.,  {Verde} L.,  {Halpern} M.,  {Hill} R.~S.,  {Kogut} A.,
  {Limon} M.,  {Meyer} S.~S.,  {Odegard} N.,  {Tucker} G.~S.,  {Weiland} J.~L.,
   {Wollack} E.,    {Wright} E.~L.,  2007, \apjs, 170, 377

\bibitem[\protect\citeauthoryear{{Springel}}{{Springel}}{2005}]{gadget2}
{Springel} V.,  2005, \mnras, 364, 1105

\bibitem[\protect\citeauthoryear{{Stark}, {Ellis}, {Chiu}, {Ouchi} \&
  {Bunker}}{{Stark} et~al.}{2010}]{stark10}
{Stark} D.~P.,  {Ellis} R.~S.,  {Chiu} K.,  {Ouchi} M.,    {Bunker} A.,  2010,
  \mnras, 408, 1628

\bibitem[\protect\citeauthoryear{{Steidel}, {Adelberger}, {Giavalisco},
  {Dickinson} \& {Pettini}}{{Steidel} et~al.}{1999}]{steidel99}
{Steidel} C.~C.,  {Adelberger} K.~L.,  {Giavalisco} M.,  {Dickinson} M.,
  {Pettini} M.,  1999, \apj, 519, 1

\bibitem[\protect\citeauthoryear{{Tapken}, {Appenzeller}, {Gabasch}, {Heidt},
  {Hopp}, {Bender}, {Mehlert}, {Noll}, {Seitz} \& {Seifert}}{{Tapken}
  et~al.}{2006}]{tapken06}
{Tapken} C.,  {Appenzeller} I.,  {Gabasch} A.,  {Heidt} J.,  {Hopp} U.,
  {Bender} R.,  {Mehlert} D.,  {Noll} S.,  {Seitz} S.,    {Seifert} W.,  2006,
  \aap, 455, 145

\bibitem[\protect\citeauthoryear{{Tapken}, {Appenzeller}, {Mehlert}, {Noll} \&
  {Richling}}{{Tapken} et~al.}{2004}]{tapken04}
{Tapken} C.,  {Appenzeller} I.,  {Mehlert} D.,  {Noll} S.,    {Richling} S.,
  2004, \aap, 416, L1

\bibitem[\protect\citeauthoryear{{Tapken}, {Appenzeller}, {Noll}, {Richling},
  {Heidt}, {Meink{\"o}hn} \& {Mehlert}}{{Tapken} et~al.}{2007}]{tapken}
{Tapken} C.,  {Appenzeller} I.,  {Noll} S.,  {Richling} S.,  {Heidt} J.,
  {Meink{\"o}hn} E.,    {Mehlert} D.,  2007, \aap, 467, 63

\bibitem[\protect\citeauthoryear{{Tenorio-Tagle}, {Silich}, {Kunth},
  {Terlevich} \& {Terlevich}}{{Tenorio-Tagle} et~al.}{1999}]{tenorio99}
{Tenorio-Tagle} G.,  {Silich} S.~A.,  {Kunth} D.,  {Terlevich} E.,
  {Terlevich} R.,  1999, \mnras, 309, 332

\bibitem[\protect\citeauthoryear{{Tilvi}, {Malhotra}, {Rhoads}, {Scannapieco},
  {Thacker}, {Iliev} \& {Mellema}}{{Tilvi} et~al.}{2009}]{tilvi09}
{Tilvi} V.,  {Malhotra} S.,  {Rhoads} J.~E.,  {Scannapieco} E.,  {Thacker}
  R.~J.,  {Iliev} I.~T.,    {Mellema} G.,  2009, \apj, 704, 724

\bibitem[\protect\citeauthoryear{{Tweed}, {Devriendt}, {Blaizot}, {Colombi} \&
  {Slyz}}{{Tweed} et~al.}{2009}]{tweed09}
{Tweed} D.,  {Devriendt} J.,  {Blaizot} J.,  {Colombi} S.,    {Slyz} A.,  2009,
  \aap, 506, 647

\bibitem[\protect\citeauthoryear{{Valls-Gabaud}}{{Valls-Gabaud}}{1993}]{valls}
{Valls-Gabaud} D.,  1993, \apj, 419, 7

\bibitem[\protect\citeauthoryear{{Van Breukelen}, {Jarvis} \& {Venemans}}{{van
  Breukelen} et~al.}{2005}]{vanb05}
{van Breukelen} C.,  {Jarvis} M.~J.,    {Venemans} B.~P.,  2005, \mnras, 359,
  895

\bibitem[\protect\citeauthoryear{{Verhamme}, {Schaerer}, {Atek} \&
  {Tapken}}{{Verhamme} et~al.}{2008}]{verh08}
{Verhamme} A.,  {Schaerer} D.,  {Atek} H.,    {Tapken} C.,  2008, \aap, 491, 89

\bibitem[\protect\citeauthoryear{{Verhamme}, {Schaerer} \&
  {Maselli}}{{Verhamme} et~al.}{2006}]{verh06}
{Verhamme} A.,  {Schaerer} D.,    {Maselli} A.,  2006, \aap, 460, 397

\bibitem[\protect\citeauthoryear{Verhamme}{{Verhamme} et~al.}{2012}]{verh12}
{Verhamme} A., {Dubois} Y., {Blaizot} J., {Garel} T., {Bacon} R.,
{Devriendt} J., {Guiderdoni} B., {Slyz} A.,  2012, \aap, page submitted

\bibitem[\protect\citeauthoryear{{Wang}, {Malhotra}, {Rhoads}, {Zhang} \&
  {Finkelstein}}{{Wang} et~al.}{2009}]{wang09}
{Wang} J.,  {Malhotra} S.,  {Rhoads} J.~E.,  {Zhang} H.,    {Finkelstein}
  S.~L.,  2009, \apj, 706, 762

\bibitem[\protect\citeauthoryear{{Zheng}, {Cen}, {Trac} \&
  {Miralda-Escud{\'e}}}{{Zheng} et~al.}{2010}]{zheng10}
{Zheng} Z.,  {Cen} R.,  {Trac} H.,    {Miralda-Escud{\'e}} J.,  2010, \apj,
  716, 574

\bibitem[\protect\citeauthoryear{{Zheng} \& {Miralda-Escud{\'e}}}{{Zheng} \&
  {Miralda-Escud{\'e}}}{2002}]{zheng02}
{Zheng} Z.,  {Miralda-Escud{\'e}} J.,  2002, \apj, 578, 33

\end{thebibliography}

\end{document}